\documentclass[10pt]{iopart} 

\usepackage{graphicx}
\usepackage{amssymb}
\usepackage{amsfonts}
\usepackage{caption}
\usepackage{multirow}
\usepackage{lipsum}
\usepackage{xcolor}
\usepackage{floatrow}
\usepackage{cite}
\usepackage{hyperref}

\usepackage{iopams}

\newcommand{\CMO}{CaMoO$_4$}
\newcommand{\CWO}{CaWO$_4$}

\newcommand{\zerodbd}{$0\nu\beta\beta$}
\newcommand{\twodbd}{$2\nu\beta\beta$}
\newcommand{\LMO}{Li$_2$MoO$_4$}
\newcommand{\enLMO}{Li$_2$$^{100}$MoO$_4$}

\newcommand{\mbb}{$ m_{\beta\beta}$}

\newcommand{\Tzerov}{$T_{1/2}^{0\nu}$}

\newcommand{\Mo}[1]{$^{#1}$Mo}

\newcommand{\Se}[1]{$^{#1}$Se}
\newcommand{\Te}[1]{$^{#1}$Te}
\newcommand{\Ca}[1]{$^{#1}$Ca}

\begin{document}

\title{Superconducting detectors for rare event searches in experimental astroparticle physics}

\author{ Yong-Hamb Kim$^{1,4,5}$, Sang-Jun Lee$^2$, Byeongsu Yang$^3$,}

\address{$^1$ Center for Underground Physics, Institute for Basic Science (IBS), Daejeon 34047, Korea}
\address{$^2$ SLAC National Accelerator Laboratory, Menlo Park, CA 94025, USA}
\address{$^3$ Department of Physics and Astronomy, Seoul National University, Seoul, 08826, Korea}

\address{$^4$ University of Science and Technology (UST), Daejeon 34113, Korea}
\address{$^5$ Korea Research Institute of Standards and Science (KRISS), Daejeon 34113, Korea}

\ead{yhk@ibs.re.kr, sangjun2@slac.stanford.edu, bsyang@snu.ac.kr}


\begin{abstract}

Superconducting detectors have become an important tool in experimental astroparticle physics, which seeks to provide a fundamental understanding of the Universe. 
In particular, such detectors have demonstrated excellent potential in two challenging research areas involving rare event search experiments, namely, the direct detection of dark matter and the search for neutrinoless double beta decay. Here, we review the superconducting detectors that have been and are planned to be used in these two categories of experiments. We first provide brief histories of the two research areas and outline their significance and challenges in astroparticle physics. Then, we present an extensive overview of various types of superconducting detectors with a focus on sensor technologies and detector physics, which are based on calorimetric measurements and heat flow in the detector components. Finally, we introduce leading experiments and discuss their future prospects for the detection of dark matter and the search for neutrinoless double beta decay employing superconducting detectors.
\end{abstract}

%
%
\submitto{\SUST}
%
\maketitle
%
\ioptwocol

\tableofcontents


\section{Introduction} 
\label{sec:intro}

During the last few decades, great progress has been achieved in low-temperature detectors (LTDs) operating at sub-Kelvin temperatures, which have played a major role in many scientific applications~\cite{enssbook,mitsuda_ltd17,pirro2017ar}. 
These LTDs utilize novel properties of materials and technologies specialized for operation at low temperatures to increase their detection sensitivity. 
In particular, superconducting materials, circuits, and devices have become the key components in many state-of-the-art LTDs. Throughout this review, we define a superconducting detector as an LTD in which superconductivity plays a critical role in detecting a signal. Moreover, we distinguish a detector from a sensor, which we consider to refer only to the sensing element of a detector. Thus, a typical superconducting detector can be said to consist of a superconducting sensor and a target (or an absorber) for the particles or interactions to be detected. It should be noted that although our definition is not clear-cut in some situations, we view it as a plausible way to cover a variety of techniques at play.

One of the most important advantages of superconducting detectors is their detection sensitivity. Detection sensitivity is a rather complex (or loosely defined) concept that encompasses the energy resolution, minimum detectable energy, time resolution, detector volume/mass, detection efficiency, etc. Unfortunately, there is no universal figure-of-merit expression for detection sensitivity that would apply to every type of superconducting sensor/detector. Instead, energy resolution is often considered the most critical characteristic for sensors and detectors and is used to directly compare different techniques. Superconducting sensors/detectors have demonstrated high energy resolution far beyond the theoretical limit of conventional semiconductor devices in any given energy region. Moreover, superconducting detectors can also be made to have a high detection efficiency comparable to that of semiconductor detectors. For this reason, superconducting detectors have replaced semiconductor detectors in many applications, especially when a stringent energy resolution requirement needs to be met.

Currently, superconducting sensors and detectors are being used over a wide energy range.
For example, in the detection of infrared (IR) and visible photons, superconducting detectors demonstrate high speed and high efficiency for the detection of single photons~\cite{lita2008oe,natarajan2012sust}.
High-performance single-photon detectors of this kind have become an essential tool in optical quantum computing and quantum communications~\cite{mirhosseini2020nature
,you2020Nanophotonics}.
For the detection of X-rays in an energy range between a few hundred eV and a few hundred keV, superconducting detectors have shown much better energy resolution than semiconductor-based detectors while also offering a high detection efficiency.
Because of their unique combination of high resolution and high efficiency, a few large-scale X-ray satellite missions are being developed based on state-of-the-art superconducting detectors~\cite{Barcons2015athena,gaskin2019lynx}. 
Superconducting detectors also provide superior performance in gamma-ray and alpha radionuclide analysis~\cite{ullom2015sust,horansky2010jap,iwkim2017}. 
In particular, superconducting detectors equipped with metal 4$\pi$ absorbers enable decay energy spectroscopy for alpha- and beta-emitting radionuclides, a new spectroscopic method for the accurate measurement of spectral shapes and activities~\cite{sjlee2010,loidl2014,ranitzsch2019jltp}. 
This new method has also been adopted to study important properties of neutrinos such as the rest mass through the end-point measurement of beta-decay spectra~\cite{gastaldo2014,holmes2015}.  

Astroparticle physics experiments searching for rare events are another field of application in which superconducting detectors play key roles. 
In this field of study, unambiguous detections of rare events would lead to groundbreaking discoveries of new particles or new physical processes. 
One such example is the direct detection of dark matter (DM), which is being pursued by a number of international projects.
DM is a type of matter that is known to exist in the Universe, as evidenced by many indirect measurements~\cite{Rubin:1970zza,Rubin:1980zd,Massey:2010hh,hinshaw2013nine,aghanim2020planck,markevitch2004aj,Allen:2011aa}.
Understanding the nature of DM is one of the most mysterious but fundamental research topics in physics. Several leading DM search projects have adopted superconducting detectors as their main detectors~\cite{Agnese:2017njq,Abdelhameed:2019hmk}.  
In particular, in experimental searches for particle-like DM, superconducting detectors are primarily intended to measure heat signals originating from the energy deposited in a target/absorber material when a DM particle interacts with normal matter in the target. 
Superconducting detectors enable the detection of heat signals with high resolution as well as a significantly lowered minimum detectable energy (i.e., energy threshold) compared with conventional detectors. To further increase the detector sensitivity to DM-normal matter interactions, an additional detection channel can be utilized, especially for a target consisting of a semiconductor or scintillating material, from which charge or light signals, respectively, can be measured together with the main heat signals. Such dual-channel detection makes it possible to distinguish whether a detected signal comes from unwanted background or indeed from the rare events being sought. Particle identification (PID) of this kind is now an essential capability of rare event search detectors.
 
 Another example of the use of superconducting detectors in this field is the search for neutrinoless double beta decay (\zerodbd). \zerodbd{} is a hypothetical decay process that would occur if neutrinos have finite mass and are their own antiparticles (i.e., Majorana particles)~\cite{Majorana2006}. Its experimental observation would not only reveal basic but unknown characteristics of neutrinos but also provide a fundamental understanding of the origin of the present matter-dominated Universe. A number of detection technologies have been developed to probe the rare events of \zerodbd\cite{adams2020prl,agostini2020prl,gando2016prl,al2021search}.
Superconducting detectors are one promising detection method because of their energy resolution, efficiency and PID capability.

Here, we review superconducting detectors in the context of their applicability for rare event search experiments in astroparticle physics. In Section~\ref{sec:challenges}, we introduce the significance and challenges of two categories of rare event experiments, namely, the direct detection of particle-like/wave-like DM and the search for \zerodbd. In Section~\ref{sec:sensor}, we present the basic detection principles of state-of-the-art superconducting sensors. This sensor technology section includes a discussion on low-temperature heat transfer processes as the essential mechanisms of particle detection in superconducting detectors. In Section~\ref{sec:app}, we survey direct DM detection and \zerodbd{}  search experiments based on superconducting detectors as well as other types of LTDs. This application section includes new approaches based on superconducting sensors coupled to low-temperature cavities and resonators targeted at detecting wave-like DM. We also introduce newly developed nanometer-scale superconducting detectors targeted at detecting ultralow-mass particle-like DM.

\section{Astroparticle Physics Challenges}
\label{sec:challenges}

As introduced in the previous section, the two main applications of superconducting detectors addressed in this review are direct DM detection experiments and \zerodbd{} search experiments.
These are two of the most rewarding but challenging topics in contemporary astroparticle physics. Moreover, they are the two main research areas that originally stimulated the intensive development of then-new LTD technologies. 
Thus, it is timely to describe the history and current status of these two topics before we review the corresponding sensor technologies and their applications in detail in the following sections.

First, we discuss DM, in particular, various DM candidates and strategies for their direct detection. It is commonly believed that DM does, in fact, exist based on astronomical observations. 
However, its existence has not been confirmed through the direct detection of interactions between normal matter and DM. 
Thus, the direct detection of DM is the first step in understanding its properties. 
Then, we discuss neutrinos, in particular, the known and unknown properties of neutrinos as well as the hypothetical \zerodbd{} decay and its far-reaching implications in astroparticle physics.

These two areas of research are considered among the most straightforward and revealing approaches for studying DM and neutrinos, respectively. 
However, despite several decades of effort, no clear experimental demonstration has been achieved, leading to calls for more in-depth investigations, possibly involving entirely new detector techniques. 
Superconducting detectors and related techniques are expected to play a major role in overcoming these challenges by virtue of their excellent energy sensitivity, detection efficiency, and PID capability.

\begin{figure}[t]
\includegraphics[width=7cm]{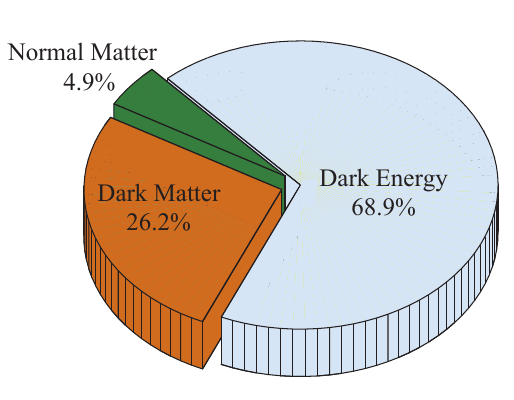}
\caption{
The proportions of normal matter, DM and dark energy in the Universe as measured by the Planck spacecraft~\cite{aghanim2020planck}.
}
\label{fig:ocupa}
\end{figure}

\subsection{Dark matter}

According to recent precise astronomical measurements  of the cosmic microwave background (CMB), normal (or ordinary) matter such as the protons, neutrons and atoms composing stars and galaxies has been found to represent only 4.9\% of the entire contents of the Universe~\cite{aghanim2020planck}. This observation also shows that another type of matter called \textit{dark matter} (DM) corresponds to approximately 26.2\% of the mass of the Universe. DM does not appear to undergo electromagnetic interactions but does interact via gravity. The most abundant component, at 68.9\%, is called dark energy, which is uniformly spread throughout the Universe and is responsible for the repulsive force that is accelerating the expansion of the Universe. The fractional contents of normal matter, DM, and dark energy in the Universe are illustrated in Figure~\ref{fig:ocupa}.

Although DM remains mysterious, our understanding of it has evolved with the advancement of technology in physics and astronomy. 
According to a recent review of the history of DM research~\cite{Bertone:2016nfn}, early scientific discussions on DM began among astronomers in the 19th century as new astronomical observations became available. In the early 20th century, some quantitative estimations of DM abundance were made with newer observations such as the velocity dispersion of the stars in the Milky Way galaxy~\cite{Bertone:2016nfn} or the motion of galaxies in a galaxy cluster~\cite{Zwicky:1933gu,zwicky1933pr}.

\begin{figure*}[t] 
\includegraphics[width=10cm]{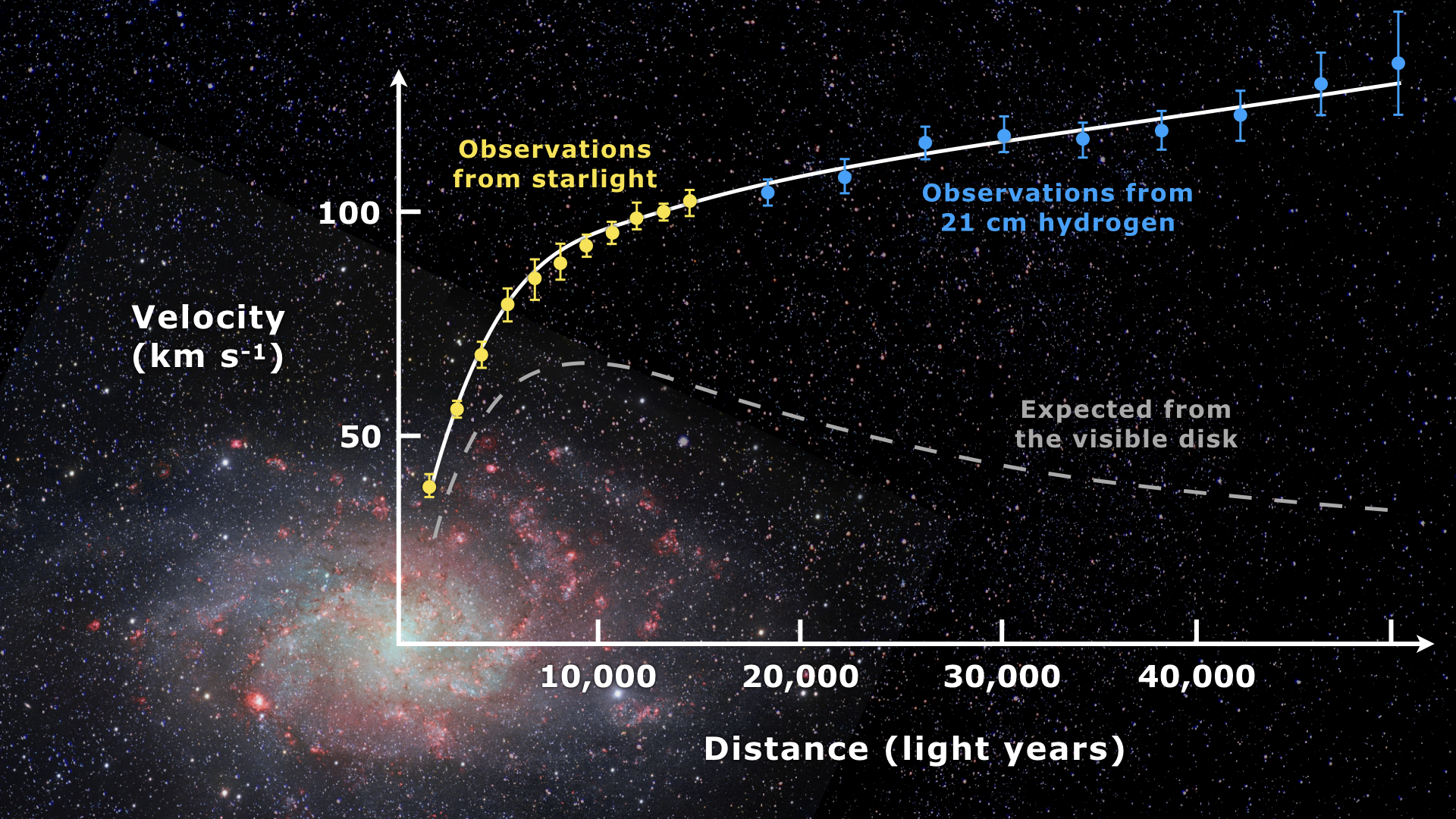} 
\caption{
An example of the rotation curve measured from the spiral galaxy Messier 33~\cite{Corbelli:2000}. The data points indicate observations of the speeds of stars as a function of the distance from the galactic center. The dashed line is the curve that is expected from the visible matter in the galaxy. The origin of the radial distance represented on the \textit{x}-axis is located at the center of the galaxy (figure from \cite{rotation_curve}).
}  
\label{fig:rot_curve}
\end{figure*}

By the 1980s, the existence of DM in some form had become widely accepted. This was mainly due to compelling observations of the velocity distributions of stars, gas and dust in galaxies with respect to the distances from the centers of these galaxies reported by Vera Rubin, Kent Ford and Ken Freeman~\cite{Rubin:1970zza,Rubin:1980zd,Freeman:1970mx}. 
Figure~\ref{fig:rot_curve} shows an example of the rotation curve of the stars in a spiral galaxy. From the visible components, the rotation curve is expected to slow down for stars in the outer spirals of the galaxy. However, the observations show that stars far from the galactic center rotate much faster than expected. This indicates that the amount of visible normal matter is not sufficient to explain the measured rotation curve. Consequently, some unobserved origin of gravitation (i.e., DM) must exist to hold the stars tightly within the galaxy. 
Observational evidence from a number of other sources has been reported that also indicates the existence of DM. Those sources include gravitational lensing~\cite{Massey:2010hh}, the CMB \cite{hinshaw2013nine,aghanim2020planck}, the Bullet Cluster~\cite{markevitch2004aj}, the total masses of galaxy clusters~\cite{Allen:2011aa}, and structure formation in the early Universe~\cite{Primack:1997av}.

\begin{figure*}[t]
\centering
\includegraphics[width=16cm]{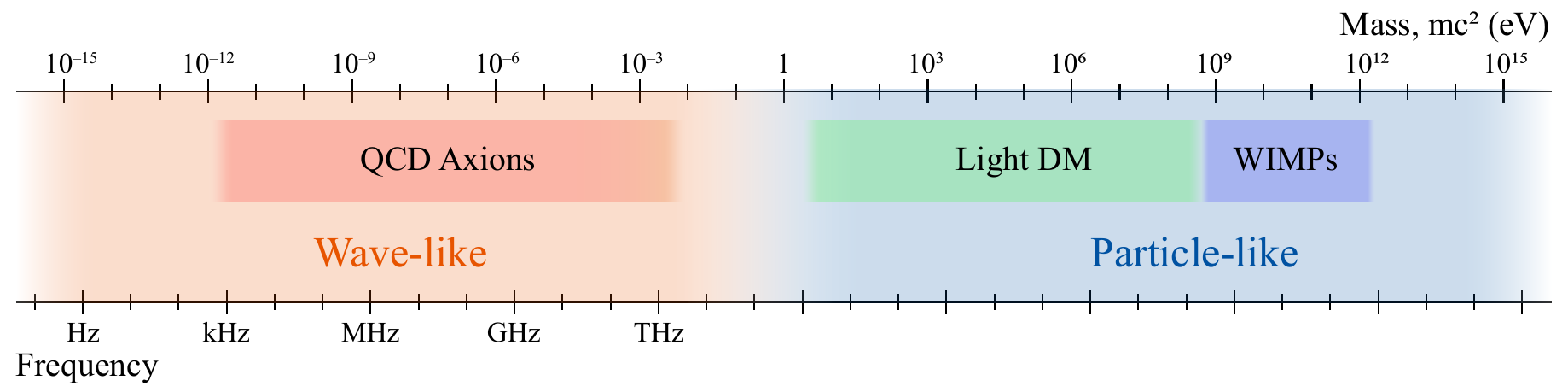}
\caption{
Three strong DM candidates, namely, axions, light DM and WIMPs, and their expected mass ranges. When the mass of a DM particle is smaller (greater) than $\sim 0.1$\,eV,  that particle shows wave-like (particle-like) characteristics. Many other candidates that are not listed also exist; some of them even have expected mass ranges outside of the mass range shown here~\cite{battaglieri2017}.
}
\label{fig:dm_candidate}
\end{figure*}

As discussed above, scientists have found evidence for the existence of DM mostly from its gravitational interactions with normal matter. However, almost no observations are available to explain the nature or properties of DM. The most pressing fundamental question about DM is ``\textit{What is DM}?''  It is a rather embarrassing question, but it accurately represents the current status of our understanding about DM: many theoretical postulates exist, but none of them has been proven to be the ultimate explanation. Following from this, ``\textit{Can we detect DM}?''  is another question of this kind. It is like looking for something without knowing if it even exists. Thus, any experimental hint of the direct interaction of DM with normal matter might be the first step towards answering these questions. Such DM-normal matter interactions may involve the generation, decay, or annihilation of DM.

Here, we briefly describe our present understanding of DM. Although the mass-energy density of DM as a whole has been estimated  
with a relatively high accuracy, its individual mass is a largely unknown parameter. There is a vast possible range, between 10$^{-22}$\,eV and 10$^{67}$\,eV, set by various cosmological constraints~\cite{Hlozek:2014lca,Armengaud:2017nkf,Nori:2018pka,Nadler:2019zrb,monroy2014end,Brandt:2016aco}. 
Note that eV is used here as a unit of mass that corresponds to a mass of eV$/c^2$, where $c$ is the speed of the light. Figure~\ref{fig:dm_candidate}  illustrates a partial region of the DM mass range that includes the expected mass ranges of three strong DM candidates, namely, axions, light DM, and weakly interacting massive particles (WIMPs). These three candidates are further discussed below. Because of the wide range of possible DM masses, very different properties of DM are expected depending on its mass scale. For instance, if the mass were to fall in the range of $10^{-22}\,\mathrm{eV} < m_\mathrm{DM} < 0.1\,\mathrm{eV}$, the DM would behave like waves with a de\,Broglie frequency $f$ determined by $h f = m_\mathrm{DM} c^2$,  where $h$ is the Planck constant and $m_\mathrm{DM}$  is the DM rest mass. Thus, experiments targeting the detection of DM in this mass range should be sensitive to signals originating from the wave-like properties of the DM. Axions are a wave-like DM candidate. On the other hand, for DM in the mass range of 
$0.1\,\mathrm{eV} < m_\mathrm{DM} <$  $10^{19}\,\mathrm{GeV}$ ($\sim$the Planck scale), 
its prominent properties would be those of
a particle. Therefore, any corresponding detection method should utilize its particle-like behavior. WIMPs and light DM are particle-like DM candidates.
There could also exist intermediate-mass DM exhibiting the dual properties of waves and particles in the intermediate mass region~\cite{Jaeckel:2010ni,Zeldovich:1967lct,Hawking:1971ei,monroy2014end,Brandt:2016aco}.
At a much higher mass scale than the Planck scale, massive astrophysical compact halo objects (MACHOs) are also considered DM candidates, including primordial black holes (PBHs), a hypothetical species that formed soon after the Big Bang~\cite{battaglieri2017}. Below, we further describe the three strong DM candidates, namely, WIMPs, light DM, and axions.

\begin{itemize}
\item{WIMPs}

 As the evidence for the existence of DM has accumulated, many theoretical models have been suggested to explain what the DM is. At the minimum, these models should satisfy all constraints obtained from indirect observations of DM. 
For example, the DM must be electrically neutral and stable against decay, with a lifetime longer than the age of the Universe. 
Since none of the Standard Model particles meets all necessary criteria to be the DM particle, many hypothetical particles have been proposed based on various models. Among them, WIMPs have been the most favored candidate.

On the one hand, WIMPs are favored in the context of the thermodynamic evolution of the early Universe according to Big Bang cosmology. If they have a mass of approximately 100\,GeV and a cross section at the electroweak scale, they well explain the cosmological abundance of DM. On the other hand, WIMPs are also favored by a beyond-the-Standard-Model theory called supersymmetry (SUSY). SUSY postulates that \textit{super}partners exist for all the Standard Model particles (normal matter) and that they were created in great abundance in the early Universe. Among the SUSY particles, the lightest one is expected to have a mass in the range of 100\,GeV to a few TeV and to be stable enough to have persisted and constitute the present DM. In particular, their interactions with normal matter are also at the electroweak scale~\cite{Pagels:1981ke,Goldberg:1983nd,Ellis:1983ew,Jungman:1995df,Steffen:2008qp,Bertone:2004pz}. This accidental coincidence in mass and cross section between the two hypotheses is known as the \textit{WIMP miracle}. Without any better alternative, WIMPs have naturally become the most preferred and the most sought-after DM candidate.

Searching for WIMPs has been attempted in various ways. The Large Hadron Collider (LHC), with its energy within the reach of the SUSY-predicted WIMP mass, enabled a search for WIMPs that might have been created inside the collider~\cite{Kane:2008gb}. If WIMPs were indeed created by the colliding beams of protons, the WIMPs would have left the detector undetected, resulting in a so-called missing-momentum signal. However, no such signal has been detected~\cite{ATLAS:2018iyk}.

A large number of underground experiments have also searched for signals originating from the direct interaction of WIMPs with the detectors. Although most of these experiments use distinct detection techniques, including superconducting sensors, they share many common features. They all rely on the particle-like nature of WIMPs, use ultrapure materials with low radioactivities for the detectors and their supporting structures, and are located in deep underground laboratories with heavy radiation shielding to minimize any WIMP-mimicking background signals~\cite{Abdelhameed:2019hmk,Agnese:2018gze,Aprile:2018dbl}.

Figure~\ref{fig:dm_limit} shows the DM limit curves for spin-independent (SI) interactions between DM and normal matter (nucleon) obtained by a selected list of underground experiments. Among all WIMP search experiments, including those in this figure, the DArk MAtter/Large sodium Iodide Bulk for RAre processes (DAMA/LIBRA) result is the only positive claim for WIMP detection at the present time. Their claim comes from an annual modulation of the signal counts in their WIMP signal window that has been observed by their NaI crystal detectors for more than 13 annual cycles~\cite{bernabei2013epjc}. Their result is solid and compatible with the WIMP masses predicted in the ``traditional'' WIMP DM scenario. However, their claim has not been confirmed by any other experiment. Early on, various LTD experiments, including those with superconducting detectors, namely, Exp\'erience pour D\'etecter Les WIMPs En Site Souterraine (EDELWEISS)~\cite{benoit2002improved}, the Cryogenic Dark Matter Search (CDMS)~\cite{akerib2004first}, and the Cryogenic Rare Event Search with Superconducting Thermometers (CRESST)~\cite{angloher2005limits}, explored the DAMA/LIBRA region, but null results were found. Many other experiments as well as the successors of these three LTD experiments have also reported no sign of WIMP signals from various target materials. In particular, experiments with liquid xenon targets, XENON1T~\cite{Aprile:2018dbl}, the Large Underground Xenon (LUX) experiment~\cite{Akerib:2017kat}, and the Particle and Astrophysical Xenon Detector (PandaX)-II~\cite{Cui:2017nnn}, showed detection sensitivities orders of magnitude higher than the cross section predicted by the SUSY model for traditional WIMP masses, but none of them detected a positive WIMP signal. In addition, another experiment called COSINE-100 that uses the same type of NaI crystals as DAMA/LIBRA also reported no WIMP signal in the DAMA/LIBRA region~\cite{adhikari2018nature}.

Due to the null results from the LHC and several underground experiments, the DM model that was once thought to be the WIMP miracle has become highly unlikely. However, this does not mean that WIMPs are no longer a DM candidate. Rather, it means that the traditional WIMP predicted by a specific SUSY model appears to have been ruled out, whereas other types of WIMPs based on other SUSY models~\cite{olive2010pdm} as well as non-SUSY models~\cite{servant2010pdm}  are still significant DM candidates. As previously overlooked parameter space for DM is taken into consideration, many experiments that were originally targeted towards the traditional WIMP continue to improve their detection sensitivity in a wider range of WIMP masses~\cite{battaglieri2017}.

\begin{figure}[!t] 
\includegraphics[width=8cm]{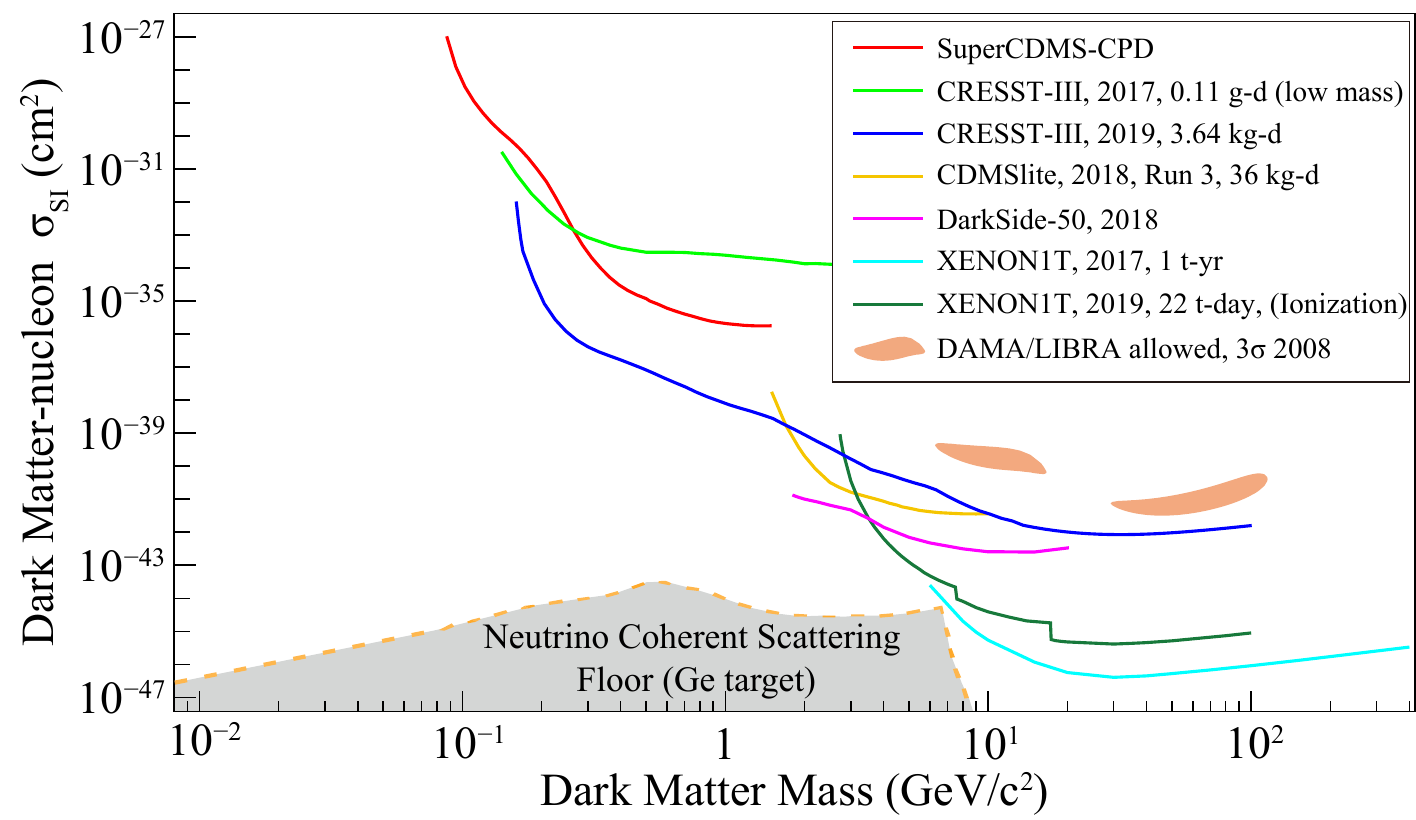} 
\caption{
DM direct detection experiments and their sensitivity limits for SI interactions between DM and nucleons. The solid curves are from the experiments that have resulted in the most stringent limits for certain mass ranges \cite{Abdelhameed:2019hmk,Alkhatib:2020slm,Agnese:2018gze,Angloher:2017sxg,Aprile:2019xxb,Aprile:2018dbl,Agnes:2018ves}. The parameter space above these curves is ruled out by the null results of the corresponding experiments. The two closed regions represent the positive claim by DAMA/LIBRA. The gray region delineated by the orange dashed curve marks the parameter space where inevitable backgrounds from coherent neutrino scattering are expected to overwhelm the DM-nucleon scattering signals for the cases of Ge targets~\cite{billard2021direct}. 
The DM sensitivity limits were extracted in March 2021 from Dark Matter Limit Plotter~\cite{dark_matter_database}.
}
\label{fig:dm_limit}
\end{figure}

\item{Light DM}

The absence of a significant signal strength for the traditional WIMP has motivated new types of DM candidates based on several different theoretical models~\cite{Duan:2017ucw,Kaplan:1991ah,Kaplan:2009ag,Petraki:2013wwa,Feng:2008ya}. There are several promising alternatives to the traditional WIMP with lighter masses and particle-like characteristics. In this review, we collectively call these alternatives light DM. Note that there is no clear-cut distinction between light DM and WIMPs. For instance, a WIMP with a lower mass than the originally expected range would be a light DM candidate~\cite{Duan:2017ucw,Gelmini:2016emn}.

An interesting light DM candidate is a sterile neutrino~\cite{adhikari2017jcap}. The sterile neutrino is a hypothetical particle whose existence was hinted at by neutrino oscillation experimental results unexpected from the three-flavor model of ordinary neutrinos~\cite{aguilar2001prd,aguilar2018prl}. Note that ordinary neutrinos are briefly discussed in the following subsection. If sterile neutrinos exist with a very weak interaction strength with normal matter and their mass is on the keV scale, they are a viable DM candidate~\cite{boyarsky2019ppnp}.

Direct detection of light DM requires a different approach from that of WIMPs. 
When a DM particle of mass $m_\mathrm{DM}$ elastically scatters off of a normal matter of mass $M$ at rest, the scattering causes the normal matter to recoil with a kinetic energy $E$:
\begin{equation}
E = \frac{\mu^2 v^2}{M} (1 -\cos \theta),
\label{eq:recoil_energy}
\end{equation}
where $\mu = m_\mathrm{DM} M/(m_\mathrm{DM}+M)$ is the reduced mass, $\boldsymbol{v}$ is the relative velocity between the DM and the normal matter, and $\theta$ is the scattering angle. 
In the case of light DM, where $m_\mathrm{DM} \ll M $, the maximum recoil energy is approximated as $2 m_\mathrm{DM}^2 v^2 / M$.
Obviously, this is much smaller than the recoil energy in the traditional WIMP case, where $m_\mathrm{DM}$ and $M$ are of the same order. Hence, the energy threshold of the detector should be significantly lowered in order to measure such a small recoil energy. 

Moreover, recent theoretical and experimental studies also consider DM interacting directly with a bound electron in a target material~\cite{battaglieri2017,essig2016jhep}.
Since the energy transferred to the rest of the atom is negligible in this process, the energy transferred to the electron $E_e$ can be approximated as the kinetic energy loss of the DM, which can be written as $\boldsymbol{q} \cdot \boldsymbol{v}-q^2/2\mu$, where $\boldsymbol{q}$ is the momentum change of the DM, and $\boldsymbol{v}$ and $\mu$ are as defined previously. Then, maximizing $E_e$ with respect to $\boldsymbol{q}$ leads to the following relation:
\begin{equation}
E_e \lesssim \frac{1}{2}\mu v^2 \sim \frac{1}{2} \,\mathrm{eV} \times \left(\frac{m_\mathrm{DM}}{\mathrm{MeV}}\right).
\label{eq:electron_recoil}
\end{equation}
This relation implies that a light DM in the MeV mass range would cause an eV-scale electron excitation, which can be measured by various high sensitivity detectors. 

For DM with a much smaller mass (sub-eV), however, the signal induced by a single DM-normal matter scattering becomes too weak (sub-$\mu$eV) to be measured by any existing detector. 
On the other hand, for the direct detection of such light-mass DMs, a totally different DM-normal matter interaction can be considered. In this scenario, a DM particle is absorbed in a superconductor via an interaction with an electron, where the DM mass converts to the energy of the electron and the phonon, in analogous to photon absorption~\cite{hochberg2016prd}. 
A DM candidate called dark photon may undergo such a hypothetical DM absorption process. A DM absorption by a electron generates quasi-particles or non-equilibrium phonons of the order meV, which can be measured with superconducting detectors such as those based on superconducting nanowires ~\cite{hochberg2016prd,hochberg2021arxiv}.
Other approaches for detecting dark photons include using an optical haloscope~\cite{chiles2021first}, superconducting qubits~\cite{dixit2021searching}, and superconducting radiofrequency cavities~\cite{romanenko2020three}.

Although in general the small mass of light DM imposes experimental challenges, it has a positive effect; 
the interaction rate of the DM-normal matter is proportional to the number density of DM, which is inversely proportional to $m_\mathrm{DM}$ because the DM abundance is rather fixed. It implies that a detector with a sufficiently low threshold has a higher chance to probe the light DM as its mass is smaller. 
This is where superconducting detectors can play a major role by virtue of their excellent energy resolution and low energy thresholds. 
As seen from the DM limit curves in Figure~\ref{fig:dm_limit}, the best limits in the low-mass region below a few keV are set by the results from SuperCDMS with a Cryogenic PhotoDetector (SuperCDMS CPD), CRESST-III, and CDMSlite. Also, more and more superconducting detectors are being developed to probe even lower mass range.

\begin{figure}[t]
\includegraphics[width=8cm]{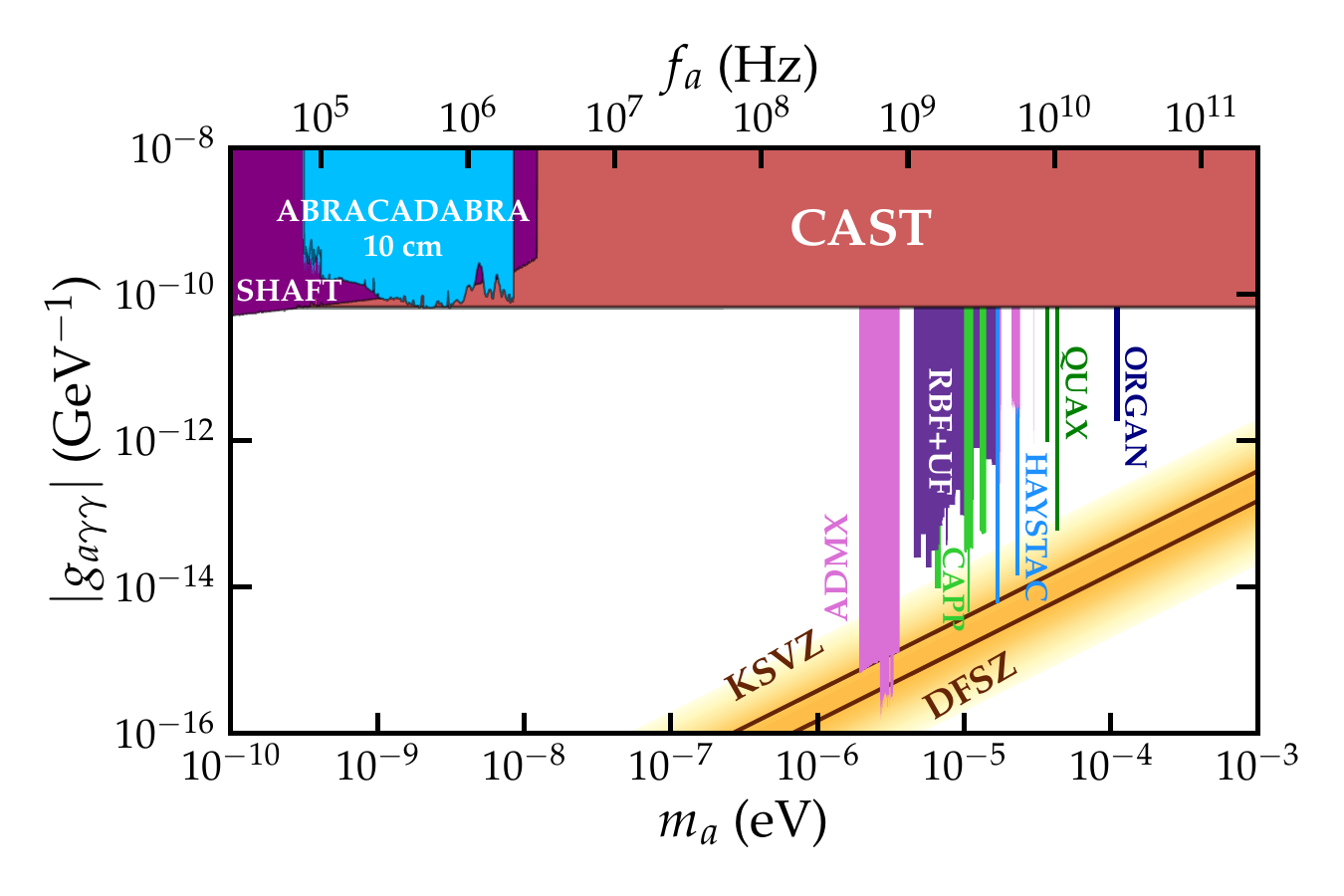}
\caption{
Experimental limits on the axion-photon coupling in the axion (also ALPs) mass range between 0.1\,neV and 1\,meV (or the frequency range between 24.18\,kHz and 241.8\,GHz). The shaded regions indicate the parameter space ruled out by the null results of axion haloscopes \cite{Asztalos:2010ADMX,Du:2018ADMX,Boutan:2018uoc,Braine:ADMX2020,
DePanfilis:1987,Hagmann:1990,Backes:2020ajv,
Brubaker:2017HAYSTAC,Lee:2020CAPP8TB,Jeong:2020cwz,Kwon:2020sav,
Alesini:2019ajt,Alesini:2020vny,McAllister:2017lkb,Gramolin:2020ict,Salemi:2021gck} and helioscopes~\cite{Anastassopoulos:2017ftl}. The two diagonal lines with orange shading represent the theoretical expectations from the KSVZ~\cite{Kim:1979if,Shifman:1979if} and DFSZ \cite{Dine:1981rt,Zhitnitsky:1980tq}  models and their variants. This figure was plotted using Ref.~\cite{ax_limit_plot}.
}
\label{fig:ax_limit}
\end{figure}

\item{Axion and axion-like particles}

Another strong candidate for the DM is \textit{axions}. Axions are hypothetical particles that were originally proposed to solve the so-called strong CP problem in quantum chromodynamics (QCD). In particle physics, the charge (C) and parity (P) symmetries determine whether the laws of physics remain the same when a particle is interchanged with its antiparticle and when its spatial coordinates are inverted to a mirror image, respectively. While the combination of the two, i.e., CP symmetry is known to be violated in electroweak interactions~\cite{christenson1964prl,alavi1999prl,aubert2001prl}, no sign of CP violation has been experimentally found in strong interactions, at least at an appreciable strength, although CP is not an \textit{a priori} conserved symmetry in QCD. This puzzle is called the strong CP problem. In 1977, as a solution to the strong CP problem, Roberto Peccei and Helen Quinn proposed a mechanism in which CP-violating terms are suppressed by a new global symmetry~\cite{Peccei:1977hh,Peccei:1977ur}. Based on this solution, Frank Wilczek and Steven Weinberg separately proposed a new type of particle~\cite{Wilczek:1977pj,Weinberg:1977ma}. Since this particle could ``clean up'' the strong CP problem, Wilczek named this particle the \textit{axion} after a laundry detergent~\cite{lamoreaux2006first}.

If axions exist, they would not only solve the strong CP problem but also be a promising DM candidate. However, the axion as proposed by Peccei, Quinn, Wilczek, and Weinberg has been experimentally ruled out~\cite{kim1987light}. Nevertheless, new theoretical mechanisms have emerged such that axions can have much weaker coupling than the original version, giving rise to so-called invisible axions. The two most viable models for the invisible axion are the Kim-Shifman-Vainshtein-Zakharov (KSVZ) \cite{Kim:1979if,Shifman:1979if} and Dine-Fischler-Srednicki-Zhitnitsky (DFSZ) models \cite{Dine:1981rt,Zhitnitsky:1980tq}. According to these two models, the axion coupling parameters are proportional to the axion mass, as plotted in Figure~\ref{fig:ax_limit}. Since the axion was conceived to have a very weak coupling, their mass is expected to be very small, and thus, they are expected to exhibit wave-like behavior, as indicated in Figure~\ref{fig:dm_candidate}. Consequently, their direct detection is expected to be extremely challenging and calls for a new approach.

In 1983, Pierre Sikivie pointed out that the axion proposed in these models would be converted into detectable photons in a strong magnetic field~\cite{1984:Sikivie}.  Based on this idea, he also proposed novel methods of building axion haloscopes and axion helioscopes to detect cosmological axions and solar axions, respectively, in a strong magnetic field. Sikivie's proposals motivated a number of axion search experiments, as represented in Figure~\ref{fig:ax_limit}. To date, however, no experiment has detected axions.

Axion DM is currently being sought using axion haloscopes, which attempt to detect photons converted from axions inside the detector by a strong applied magnetic field. Since the photon signal is expected to be extremely weak and its frequency is unknown, it is critical to build a low-noise detector with extreme sensitivity. In leading axion haloscope experiments, superconducting sensors such as superconducting Quantum interference devices (SQUIDs) and Josephson parametric amplifiers (JPAs) approaching the quantum limit are being developed as key detector components~\cite{o2020microstrip,Kutlu:2021iin}.

Experiments have also been performed to search for axion-like particles (ALPs). As the name suggests, ALPs are similar to axions, but there is one major difference. In contrast to axions, whose mass and coupling parameter are proportional, the masses and coupling parameters of ALPs are independent of each other. Many theoretical models~\cite{choi2020arxiv} predict ALPs, and as such, several experiments are aiming to detect them by scanning a wider parameter space than is addressed in axion search experiments, particularly in the low-mass region, as shown in Figure~\ref{fig:ax_limit}.

\end{itemize}


\subsection{Neutrinos and neutrinoless double beta decay}
\label{subsec:dbd_intro}

The neutrino is a subatomic particle that is abundant in the Universe. 
Its existence was first suggested by Pauli in 1932 to explain the apparent violation of energy and momentum conservation observed in the beta decay of a neutron. 
The theory of neutrinos was further developed by Fermi in 1934, and the neutrino was first experimentally discovered by Cowan and Reines in 1956~\cite{Cowan:1992xc}. 
It took such a long time for neutrinos to be directly detected because they have very little interaction with matter via the so-called weak interaction.

According to the Standard Model of particle physics, neutrinos come in three different flavor states, namely, the electron neutrino ($\nu_e$), the muon neutrino ($\nu_\mu$), and the tau neutrino ($\nu_\tau$), and they have long been considered massless. However, since the late 1990s, a number of experiments have shown compelling evidence that neutrinos convert from one flavor to another, thus establishing that neutrinos have nonzero masses and that each flavor is a different combination of three mass eigenstates ($m_i$, $i=1,2,3$)~\cite{fukuda1998prl,cleveland1998aj,ahmed2001prl,fogli2012prd}. These so-called neutrino oscillation experiments have led to measurements of the differences in the squares of the mass eigenvalues ($\Delta m^2_{ij}$), but their absolute mass scale is still unknown.

\zerodbd{} is a hypothetical decay process that can illuminate several unknown key properties of neutrinos, including the absolute mass scale. \zerodbd{} can be thought of as a special case of double beta decay. In a normal double beta decay (\twodbd), two electrons ($2\beta^-$) and two antineutrinos ($2 \bar \nu_e$) are emitted, expressed as
\begin{equation}
(Z,A)\rightarrow(Z+2,A)+2 \beta^- + 2 \bar \nu_e,
\end{equation}  
where ($Z$,$A$) stands for a nucleus with atomic number $Z$ and mass number $A$.
This process, first proposed in 1935~\cite{goeppert1935}, is now a well-established second-order weak process and has been observed in many isotopes, 
although with an extremely long half-life on the order of  $T^{2\nu}_{1/2} \approx 10^{19}$--$10^{24}$\,years~\cite{barabash2015}. 
In the meantime, double beta decay that is not accompanied by the emission of two antineutrinos was independently proposed by E. Majorana and G. Racah~\cite{majorana1937,racah1937} and further detailed by W.H. Furry~\cite{furry1939}, expressed as
\begin{equation}
(Z,A)\rightarrow(Z+2,A)+2 \beta^-.
\label{equation_0nbb}  
\end{equation}
This process can occur only if neutrinos are Majorana particles (i.e., a neutrino's antiparticle is identical to itself). Hence, the unambiguous observation of \zerodbd{} would confirm that neutrinos are Majorana and not Dirac particles, thereby settling a major unknown about neutrinos~\cite{avignone2008rmp}. Furthermore, its observation would lead to another profound consequence: it would indicate that lepton number is not always conserved, as the lepton numbers on either side of Eq.~\ref{equation_0nbb} differ by 2. The discovery of lepton number violation would have far-reaching implications in cosmology as well as in particle physics~\cite{chun2018probing}.

\begin{figure}[t] 
\begin{center}
\includegraphics[width=7cm]{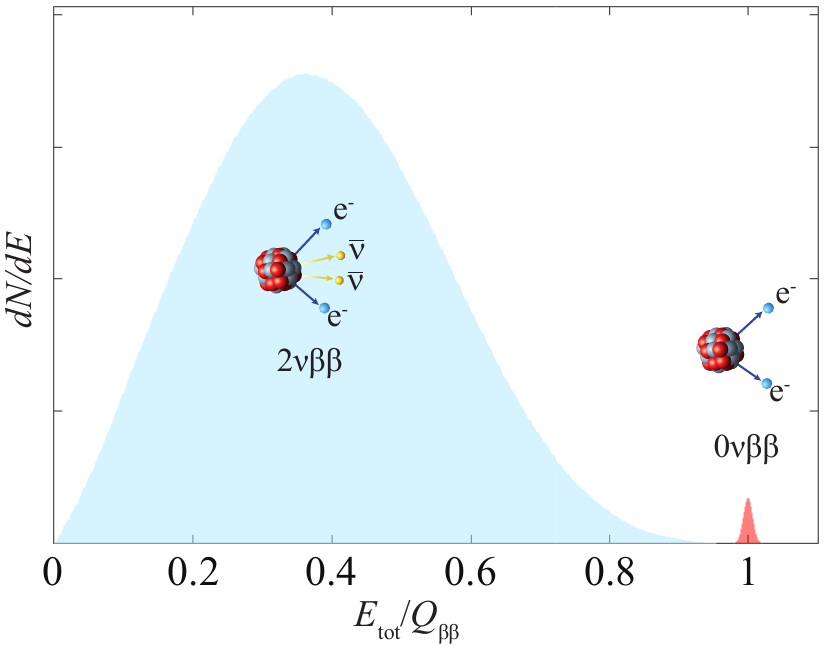}
\end{center}
\caption{
Energy spectra of the two electrons from the \twodbd{} and \zerodbd{} processes. The \zerodbd{} rate is highly exaggerated for visibility. Backgrounds and statistical fluctuations are not included in these spectra.
}
\label{dbd_spec}
\end{figure}

Experimentally, there is a large difference in the energy spectra of \twodbd{} and \zerodbd. Since the energy carried by neutrinos cannot be measured in a regular experimental setting and the recoil energy of the daughter nucleus ($Z +2, A$) is negligible, only the energy spectrum of the two electrons emitted from the decay is experimentally relevant. In the \twodbd{} case, the energy carried by the two electrons forms a broad spectrum with its end point equal to the Q-value of the decay, as shown in Figure~\ref{dbd_spec}. On the other hand, since the total decay energy is split only between the two emitted electrons in the \zerodbd{}  case, the corresponding energy spectrum is simply a peak at the Q-value.

Without any further interpretation, clear detection of the \zerodbd{} peak itself would already reveal two important conclusions: that neutrinos are Majorana-type particles and that lepton number is not strictly conserved. Even more valuable information could be obtained through a detailed quantitative analysis of a detected \zerodbd{} peak. In the case of the light neutrino exchange model, which is a commonly accepted theoretical mechanism, the decay rate $ \Gamma_{0\nu}$ of the \zerodbd{} process that would be measured is expressed as
\begin{equation}
        \Gamma^{0\nu} = 1/ T^{0\nu}_{1/2}   = G_{0\nu} \left| M_{0\nu} \right|^2  m_{\beta\beta}^2,
\label{equation_decayrate}    
\end{equation}
where $T^{0\nu}_{1/2}$ is the \zerodbd{} half-life of the element; $G_{0\nu}$ and $M_{0\nu}$ are the phase space factor and nuclear matrix element (NME), respectively, of the decay; and \mbb{}  is the effective Majorana mass. Although the phase space factor $G_{0\nu}$ can be rather accurately calculated in the framework of atomic physics~\cite{kotila2012prc}, the NME is subject to model-to-model variations by up to a factor of 3, thus presenting a need for more accurate calculations~\cite{engel2017rpp}. 
Moreover, in addition to the light neutrino exchange mechanism, there are also other theoretical models that lead to \zerodbd{} rates of the same order as that from the light neutrino exchange model~\cite{deppisch2012jpg}. Consequently, due to the current theoretical uncertainties, even a precisely measured \zerodbd{} decay rate would not translate into a well-defined $m_{\beta\beta}$ value or limit. Thus, \zerodbd{} should ultimately be measured using multiple isotopes. 

During the last two decades, a number of experiments have been carried out to search for the \zerodbd{} process in various candidate isotopes. Although different experiments have used different detection techniques, there are several common strategies for increasing the detection sensitivity. First, it is crucial to minimize any possible background events from the environment as well as the detector itself. This means that any \zerodbd{} search experiment should be carried out in a deep underground laboratory where the flux of cosmic muons is significantly suppressed. In addition, radiopure materials must be used as the materials that constitute and surround the detector. In particular, the amounts of impurities consisting of U and Th compounds in the detector materials should be reduced to unprecedentedly low levels. Second, the greater the exposure (i.e., the product of the amount of the $\beta\beta$ isotope and the measurement time) is, the higher the chance of detecting \zerodbd{} becomes. Moreover, in the presence of a background signal, a higher energy resolution would lower the count of background events in the region of interest (ROI), thereby improving the detection sensitivity, because the width of the ROI is determined by the detector's energy resolution. These strategies have motivated various technological developments in terms of detection methods, active and passive shielding, material purification and radioassays, laboratory controls/maintenance, and online/offline analysis.

In several \zerodbd{} search experiments, LTDs with crystal absorbers have been chosen as the main detection technique, primarily due to their high energy resolutions. It is also advantageous to use detectors of this type because they provide great flexibility in the selection of the target crystal, enabling an experiment to run with various crystals that contain different \zerodbd{}  candidate isotopes. To date, $^{48}$Ca, $^{82}$Se, $^{100}$Mo, $^{116}$Cd, and $^{130}$Te  have been tested using LTDs. 
These candidate isotopes with relatively high Q-values are especially suited for experiments using LTDs because of their availability in the crystal form rather than as a gas or liquid. 
Moreover, for many candidate isotopes, it is possible to realize event-by-event detection with two different types of signals for PID using LTDs because the phonon (heat) channel is almost always available in most materials. Active \zerodbd{}  experiments based on low-temperature measurements are reviewed in Section ~\ref{subsec:app-dbd}.


\section{Sensor Technologies}
\label{sec:sensor}

In this section, we discuss the sensor technologies used in superconducting sensors and detectors based on the low-temperature calorimetric measurement of energy, also known as superconducting microcalorimeters. Although the focus is on superconducting microcalorimeters, it should be noted that much of the discussion herein also applies to any microcalorimeters in general. Additionally, note that other types of superconducting sensors and detectors are briefly discussed in Section~\ref{sec:app}.

\subsection{Motivation}

The development of superconducting microcalorimeters is motivated by the need for energy sensitivity, i.e., a high energy resolution and a low energy threshold, beyond the level that conventional semiconductor- or scintillator-based detectors can achieve. Here, we present a brief comparison between conventional detectors and superconducting microcalorimeters.

When a particle or radiation interacts with a target (absorber) material in a detector, energy transfer occurs in the absorber.
The details of the energy transfer mechanism depend upon the type and energy scale of the radiation and the material and geometry of the absorber. 
Hence, the absorber should be properly chosen for a given type of radiation to ensure the efficient conversion of the input energy into measurable physical quantities. The microscopic interactions that occur during the energy transfer process, although rather complex and often not fully understood, are critical for achieving good detector performance.

The absorbers of conventional ionizing radiation detectors are made of gas, liquid, or semiconductor materials in which the  initial radiation creates measurable quantities of charge (electrons and ions) or light. 
For instance, in a semiconductor detector, an energy input creates electron-hole (e-h) pairs, which are then collected and measured by means of an electric field applied in the target volume.
In general, the signal becomes larger as more e-h pairs are created. Thus, the energy resolution of semiconductor detectors is confined by the Fano limit, representing the statistical fluctuation in the number of created e-h pairs~\cite{fano1947ionization}. The Fano-noise-limited energy resolution is expressed as $E_\mathrm{rms}$ = $\sqrt{ f w E}$. Here, $f$ is the material-specific Fano factor, $w$ is the average energy needed to create an e-h pair, which is about three times larger than the band gap of the semiconductor because of energy loss in the creation of phonons; and $E$ is the input energy.
   For Si-based detectors, where $w$ = 3.7\,eV and $f$ = 0.115, $E_\mathrm{rms}$ of 51\,eV (or, equivalently in full width at half maximum (FWHM),  $E_\mathrm{FWHM}$ = 120\,eV) is achievable for $E$ = 6\,keV, which corresponds to an energy resolving power ($E/E_\mathrm{FWHM}$) of $\sim 50$. For higher-energy alpha particles, Si-based detectors show $E_\mathrm{FWHM}$ = 8.5\,keV at 5.5\,MeV, corresponding to an energy resolving power of $\sim 650$~\cite{steinbauer1994nima}. Similarly, the intrinsic energy resolution limit of scintillator-based detectors also depends on the number of created e-h pairs, which generate scintillation photons by transferring their energy to the luminescent centers of the scintillator materiel. However, in general they have poorer resolution than semiconductor-based detectors due to a combination of low light yield and inefficient photon collection.

Unlike these conventional detectors, low-temperature thermal detectors (microcalorimeters) measure phonons created by the absorption of a particle or radiation in an absorber made of a condensed matter that is typically maintained at a temperature below 1\,K.
Although phonon measurement is more efficient than other detection channels because the majority of the input energy is converted into the phonon channel, achieving high-accuracy measurements in this channel is often very challenging. This is because phonons exist in every material, with a certain statistical frequency/energy distribution, and thermal fluctuations between the constituents of the detector can consequently overwhelm the phonon signal to be detected. However, at low temperatures, phonon measurement can be the most sensitive detection channel since the specific heat of most materials becomes very small, resulting in a larger temperature change for a given energy input, and thermal fluctuations are simultaneously greatly suppressed. Furthermore, at sufficiently low temperatures, superconducting materials and superconducting devices and electronics can be adopted to further improve the detector sensitivity. 

The fundamental limit of the energy resolution for such microcalorimeters is primarily set by the thermodynamic energy fluctuation between the detector and its thermal reservoir~\cite{enssbook:thermal}. In reality, however, noise from the readout electronics and the sensor itself (e.g., Johnson noise in the case of a detector with a resistive sensor) is often the primary factor limiting the energy resolution. In recent years, significant improvements have been achieved in the development of sensor materials and readout technologies, resulting in energy resolutions that are better than those of conventional detectors by orders of magnitude. For example, the highest energy resolutions achieved with superconducting microcalorimeters are a 1.3--1.6\,eV FWHM at 6\,keV~\cite{dewit2020optimizing,smith2012small,kempf2018} and a 0.9--1.1\,keV FWHM at 5.5\,MeV~\cite{iwkim2017,horansky2010jap}. Furthermore, with a carefully chosen absorber and detection scheme, the high resolution phonon measurement technique can be accompanied by measurements in other detection channels,  such as ionization and scintillation. Such simultaneous measurement schemes have become the gold standard in rare event search experiments using LTDs since they can be used to suppress background events. In the following subsections, some of the most mature and powerful superconducting sensors and their working principles will be discussed.

\subsection{Thermal calorimetric detection}

A typical microcalorimeter is composed of an absorber and a temperature sensor, both of which have weak thermal links to a heat reservoir (also called the heat bath). 
The initial interaction for the measurement of a particle or radiation occurs in the absorber in the form of energy transfer.
The temperature sensor then reads out
the temperature change in the detector caused by the energy transfer in the absorber. 
Typically, there is a weak thermal connection
between the heat bath and the temperature sensor, while the thermal connection between the absorber and the temperature sensor is much stronger. The heat bath is maintained at a temperature below 1\,K.

\begin{figure}[t] 
\begin{center}
\includegraphics[width=8cm]{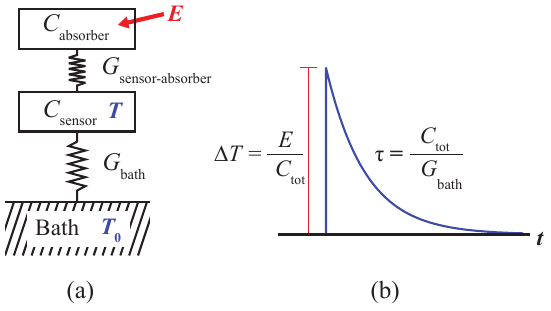}
\end{center}
\caption{(a) A simplified thermal model of a typical microcalorimeter. An input of energy ($E$) into the absorber causes a temperature increase, which is measured by the temperature sensor. $T$ and $T_0$ denote the temperatures of the sensor and the bath, respectively. (b) The expected temperature response from an energy input. $C_{\rm tot}$ represents the sum of the heat capacities of the two thermal components in the detector system, namely $C_{\rm absorber}$ and $C_{\rm sensor}$. }
\label{fig:diagram}       
\end{figure}

Figure~\ref{fig:diagram} shows a thermal model diagram and a typical temperature signal from a simple microcalorimeter.
The absorber and the temperature sensor
are connected to each other with a thermal conductance $G_{\rm sensor-absorber}$. 
The thermal conductance between the absorber and the heat bath is $G_{\rm bath}$.
Without any external energy input into the absorber or the temperature sensor, they are in thermal equilibrium at temperature $T_0$. Any energy input converted to thermal energy is eventually released into the heat bath. 

The detection principle of a microcalorimeter is straightforward. Under the assumptions that the heat transfer from the absorber to the sensor is much more efficient than that from the heat link to the bath (i.e.,  $G_{\rm sensor-absorber}$ $\gg$  $G_{\rm bath}$) and that the internal thermal conductivity of the absorber is also much larger than $G_{\rm bath}$, the temperature change $\Delta T$ in the sensor due to an energy deposition $E$ from a particle or radiation as a function of time $t$ is

\begin{equation}
\label{eq:typical}
\Delta T(t) = T-T_0 = \frac{E}{C_{\rm absorber} + C_{\rm sensor } } e^{-t/\tau} 
\end{equation}
for $  t\ge0$,  with
\begin{equation}
\tau = \frac{ C_{\rm absorber} + C_{\rm sensor } }{G_{\rm bath}},
\end{equation}
where  $C_{\rm absorber}$ and $C_{\rm sensor}$ are the heat capacities of the absorber and  the temperature sensor,
respectively. 
In this calorimetric detection scheme,  $\Delta T$ corresponds to the signal pulse height,  and  the time constant $\tau$ corresponds to the fall-time of the signal. 
This fall-time constant can be engineered to meet specific experimental requirements on the event-rate tolerance and to avoid excessive signal pile-up by fabricating the thermal link with a desired $G_{\rm bath}$. However,  $G_{\rm bath}$ should be tuned carefully because it can also affect other critical detector parameters such as energy resolution~\cite{enssbook:mmc}.

Equation~\ref{eq:typical} implies that the signal size ($\Delta T$) can be maximized by minimizing the heat capacities of the absorber and the sensor. One obvious approach is to lower the temperature because the heat capacities of many materials become very small at low temperatures.
In addition, a material with a small specific heat can be used as the absorber. For instance, a pure dielectric crystal is a good candidate material because its specific heat can be much smaller than that of a metallic or amorphous absorber at low temperatures due to the $T^3$ dependence (the Debye law). Thus, using a dielectric crystal as an absorber can make it possible to build a large-volume or large-mass detector.
In this sense, thermal calorimetric detection at mK temperatures has a great advantage because a large selection of  absorber materials are available as a target absorber.  
This is not the case for most other detection techniques.
However, it should be noted that a small specific heat should not be the only criterion for choosing an absorber material for a rare event search experiment. Additionally, other physical properties, such as the long-term stability under low temperature and high vacuum, the internal thermal conductance and the light yield (if a light signal is to be utilized) should all be considered simultaneously.

Choosing the right temperature sensor among the various available options is as important as  the selection of the absorber material. In some cases, the choice of the sensor can be made entirely independently of the choice of the absorber, but more often, they are closely related. In the following subsection, several different superconducting sensors will be reviewed to help guide the selection of sensors for rare event search experiments.

\subsection{Superconducting sensors}
\label{subsec:sensor-sc}

Most thermometers used in everyday life and for industrial/scientific purposes utilize the temperature dependence of a physical quantity. For instance, mercury thermometers work because the volume of the liquid changes with temperature. In a thermocouple gauge, the temperature dependence of the contact potential of two dissimilar metals results in a voltage signal indicative of a temperature. A platinum thermometer has a resistance that is proportional to the temperature in its working temperature range.

Similarly, temperature sensors with extreme sensitivity can be developed by measuring some temperature dependent property of the sensor material at low temperatures ($< 1$\,K). In addition, superconducting circuits and electronics can be used to maximize the sensitivity. As previously stated, in this paper, we refer to an LTD that is made of a superconducting sensor material or utilizes superconducting circuits/electronics as a superconducting detector. Although the distinction is not clear-cut, a superconducting detector can be either a thermal equilibrium detector (thermometer) or a nonequilibrium quasiparticle detector. Quasiparticles are excitations (broken Cooper pairs) in superconductors~\cite{kivelson1990bogoliubov}. A thermometer-type detector measures the temperature increase of the phonon system in a sensor material. The two most common types of thermometers are resistive thermometers and magnetic thermometers. On the other hand, a quasiparticle detector measures an excess of quasiparticles  created via energy absorption in a superconductor . 

In the case of a resistive thermometer, in which the signal readout circuit measures the change in the electrical resistance of the sensor material due to a temperature change, the resistance change caused by the initial energy $E$ input into the absorber  can be written as 
\begin{equation}
\Delta R = \left( \frac{\partial R}{\partial T} \right) \frac{E}{C_{\rm tot}},
\end{equation} 
where $C_{\rm tot}$ is the sum of the heat capacities of the detector components (primarily those of the sensor and the absorber).  Here, the temperature dependence of the resistance, $\partial R / \partial T$, is an important parameter that determines the detector sensitivity. 

In a similar way, a magnetic thermometer is another sensitive technology using a sensor material with temperature-dependent magnetization (e.g., paramagnetic Au:Er) and a superconducting circuit. A magnetic thermometer measures the change in the magnetization of the sensor material,  expressed as
\begin{equation}
\Delta M = \left( \frac{\partial M}{\partial T} \right) \frac{E}{C_{\rm tot}}.
\end{equation} 
Superconducting circuits and electronics enable accurate measurement of $\Delta M$ for extreme detector sensitivity.


In this section, four different types of LTDs will be introduced, namely, transition edge sensors (TESs), metallic magnetic calorimeters (MMCs); it is also called as magnetic microcalorimeters with the same acronym, kinetic inductance devices (KIDs), and semiconductor thermistors. TESs and MMCs are resistive and magnetic thermometers, respectively, and an KID is a quasiparticle detector. Based on our definition, TESs, MMCs, and KIDs are all superconducting sensors. Although, in general, thermistors are not superconducting sensors, they are also discussed here because they are thermal equilibrium detectors with a working principle similar to that of TESs or MMCs and are popularly used in rare event search experiments.

\subsubsection{Transition edge sensors}

TESs are among the most sensitive temperature sensors utilizing the properties of superconducting materials. A TES is a superconducting film operating at its superconducting-normal transition temperature ($T_{\rm c}$). This superconducting film often consists of a single material (an elemental superconductor), such as tungsten (W). As an alternative, a bilayer consisting of a superconductor and a noble metal, such as Mo/Au, Mo/Cu, or Ti/Au, is also popularly chosen as the superconducting film. In the DM search experiments CDMS and CRESST, W-TESs have been used, while bilayer TESs are often chosen for high-resolution X-ray, gamma-ray, and alpha spectrometers. Using W or a bilayer as the superconducting film makes it possible to tune $T_{\rm c}$. For example, the $T_{\rm c}$ of a W film varies with its crystal structure ($\alpha$-, $\beta$-, or $\gamma$-phase~\cite{colling1995low}) and the environment in which
the film was deposited. The CRESST group uses $\alpha$-W with $T_{\rm c} \sim$ 15\,mK for their
CaWO$_4$ crystals, and CDMS uses mixed-phase W with a $T_{\rm c}$ of 50--100 mK in their detectors. In the case of a bilayer TES, its Tc can be adjusted via the proximity effect by varying the thickness ratio between the two layers~\cite{martinis2000calculation}.

\begin{figure}[b] 
\begin{center}
\includegraphics[width=8cm]{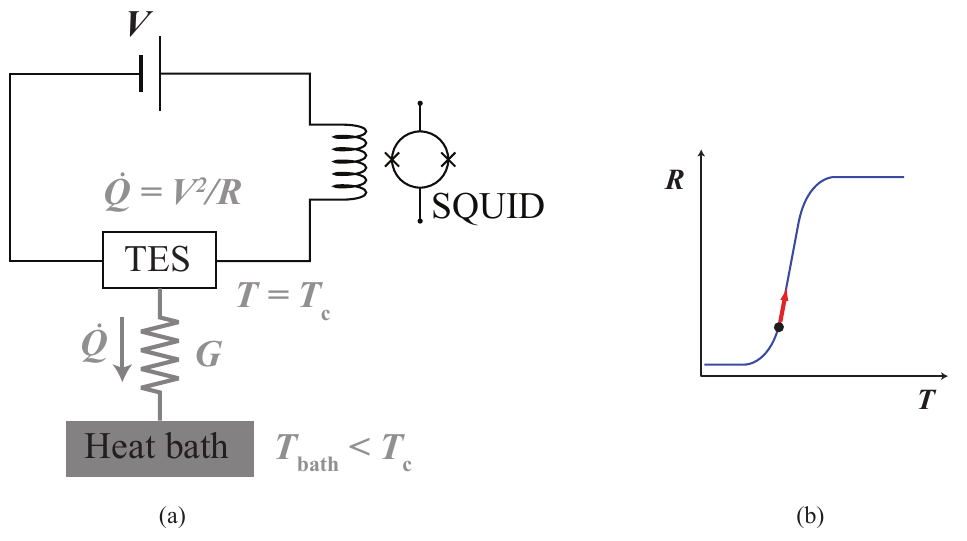}
\end{center}
\caption{
(a) Simplified electronic and thermal circuits of a TES. (b) Typical $R$-$T$ characteristics of a TES near the superconducting transition. The working point set by means of ETF is indicated by a black dot, and a temperature increase due to an energy input is indicated by a red arrow.
}
\label{fig:tes_structure}       
\end{figure}

The normal-state resistance of a TES is usually a few tens of m$\Omega$, with a transition width of a few mK or less.
Figure~\ref{fig:tes_structure} shows a typical resistance curve of a TES as a function of temperature near the superconducting-normal phase transition ``edge''. The transition width can be on the order of 0.1\,mK near a $T_{\rm c}$ of approximately 100\,mK. 
 The temperature dependence of the resistance $(\partial R/\partial T)$ is very large at the transition, making the TES a very sensitive thermometer.

TESs are, in general, operated in the voltage bias mode, such that a change in the resistance  of the TES causes a change in the  current  in the bias circuit. This current change is measured with a low-noise SQUID with high accuracy. A simplified TES geometry and the corresponding measurement circuit are illustrated in Figure~\ref{fig:tes_structure}. 
As shown in this figure, the TES film has a weak thermal connection to a heat bath that is regulated to a temperature below the $T_{\rm c}$ of the film ($T_{\rm bath} < T_{\rm c}$).
With the application of a bias voltage $V$ that is sufficiently high to break the superconductivity of the TES film, the TES becomes resistive, and the temperature of the TES is elevated from $T_{\rm bath}$ to within its superconducting-normal transition region due to Joule heating: ${P}_{\rm Joule} = V^2/R(T)$, where $R(T)$ is the resistance of the TES at $T$.
With the weak thermal link being the only thermal connection between the film and the bath, the temperature of the TES is determined by the thermal differential equation
\begin{equation} \label{eq:heat}
C \frac{dT}{dt} = - P_{\rm bath}  + P_{\rm Joule} = - P_{\rm bath}  + \frac{V^2}{R(T)},
\end{equation}
where $P_{\rm bath}$ is the power flowing from the TES to the heat bath, which is a function of  the TES temperature, the bath temperature, and the thermal conductance between the TES and the bath.

Equation~\ref{eq:heat} has important implications. When a TES is at a temperature $T_0$ within its transition region under  a constant bias voltage and there is no external energy or power input, the TES is in thermal equilibrium, with $P_{\rm bath}  = P_{\rm Joule} = V^2/R(T)$. However, once the temperature of the TES is increased (decreased) from $T_0$, the Joule heating decreases (increases) because $R(T)$ has a positive slope in the transition region. This effect is called negative electrothermal feedback (ETF). As a result, the temperature of the TES is self-regulated at the quiescent temperature $T_0$. 
The negative ETF mode has several advantages for particle detection with TESs. Because they self-regulate their temperature within the transition region, a large array of TESs in a detector can be stably operated even if they have slightly different $T_{\rm c}$ values, and they also become less sensitive to fluctuations in the bath temperature. Moreover, for a given energy input $E$, the recovery time of a TES becomes faster than its natural time constant in the absence of the ETF effect  because the excess energy is effectively removed by a reduction in the Joule heating, thereby reducing the dead time of the detector due to pulse pile-up. 

Another advantage of using a TES is that the TES film can be directly evaporated onto the surface of an absorber. Direct contact between the sensor and the absorber enables  efficient heat transfer between them, which results in a much faster response time (a rise time of $\sim$1\,ms) than those of other detectors such as thermistors ($>10$\,ms). The fast response time of TESs make them suitable for detecting athermal phonons which can help increase the detector sensitivity and facilitate the rejection of background signals.

TESs have been used as the temperature sensors  in DM search projects such as CDMS and CRESST, as will be discussed in Section~\ref{subsec:app-dm}. They have also been extensively developed for other applications, such as X-ray and nuclear spectroscopy. 
To the best of our knowledge, the best demonstrated FWHM energy resolutions are 0.7\,eV at 1.5\,keV~\cite{lee2015fine} and 1.3--1.6\,eV for 5.9\,keV X-rays~\cite{dewit2020optimizing,smith2012small}, 65\,eV  for 208\,keV gamma rays~\cite{yoho2020automated} and 1.0--1.1\,keV for 5.3\,MeV alpha particles~\cite{horansky2010jap,hoover2015measurement}. These energy resolutions are improved by orders of magnitudes compared to those of comparable commercial semiconductor detectors,  such as Si(Li) detectors and high purity Ge detectors for X-rays and gamma rays and passive implanted planar Si detectors for alpha particles.

\subsubsection{Magnetic microcalorimeters}

MMCs are the most mature magnetic thermometers used in particle detection. MMCs have become an important technology in various applications that require high energy resolution over a wide energy range~\cite{enssbook:mmc}, and they have
great potential for use in rare event search experiments with crystal absorbers~\cite{yhkim2004,sjlee2011,gbkim2017}. The best achieved energy resolution of an MMC to date is a 1.6 eV FWHM for 6 keV x-rays~\cite{kempf2018}, which is comparable to the best resolution demonstrated by a TES. In an MMC setup designed for an energy range of 3--5\,MeV, an FWHM of 0.86 keV has been obtained for the dominant Gaussian part of a 5.5\,MeV alpha signal peak~\cite{iwkim2017}. There is also another promising type of magnetic thermometer called a magnetic penetration thermometer (MPT)~\cite{nagler2012performance}, but its maturity has not reached the level of MMCs, and it will not be discussed here.

An MMC uses a paramagnetic material as the temperature sensor. This is because the magnetization of a simple paramagnetic system is inversely proportional to the temperature, following Curie's law, and the temperature sensitivity can consequently be very large at low temperatures. For many years, Au:Er, a dilute magnetic alloy of gold doped with a small concentration of erbium, has been the most popular choice of the paramagnetic sensor material for MMCs because it maintains its paramagnetic properties at temperatures of tens of mK, and the good thermal conductivity of the host metal Au guarantees fast sensor thermalization at such low temperatures. Moreover, its thermal and magnetic properties are well understood through mean field theory, which takes into account the exchange interactions between the magnetic spins in a sensor~\cite{mmc:2000}.

However, there is a drawback in using Au:Er as the sensor material of an MMC because it shows excess heat capacity arising from nuclear quadrupole moments of the gold nuclei under the electrical field gradient induced by the Er ions~\cite{enssbook:mmc,mmc:2000}. Since this excess heat capacity becomes the major contributor to the total heat capacity at temperatures below $\sim$20\,mK, Au:Er is not an ideal sensor material for rare-event search experiments with large crystal absorbers typically running at 10--20\,mK. 
As a promising alternative, Ag:Er, Er-doped silver~\cite{fleischmann2009,kempf2018}, has been quickly taking over the popularity of Au:Er. A recent study where low temperature properties of MMCs based on two types of sensor materials of Au:Er(1000 ppm) and Ag:Er(414 ppm) were measured~\cite{sgkim2021ieeetas} demonstrated the Ag:Er MMC showed no noticeable excess heat capacity and larger signal sizes than those from the Au:Er MMC.

MMCs utilize superconducting circuits and electronics to measure the changes in magnetization of the sensor material. 
Typically, superconducting coils are integrated on an MMC sensor chip to generate magnetic fields for magnetizing the sensor material and to pick up magnetization changes in the sensor. 
The signal picked up by the superconducting coils is read out by a dc-SQUID as a voltage signal.
Thus, the detection principle of an MMC can be expressed as $E$ $\rightarrow$ $\Delta T$ $\rightarrow$ $\Delta M$ $\rightarrow$ $\Delta V$, where $E$ is the input energy; $\Delta T$ and $\Delta M$ denote the change in sensor's temperature and magnetization, respectively; and  $\Delta V$ is the voltage signal of the SQUID. 

\begin{figure}[b] 
\begin{center}
\includegraphics[width=8cm]{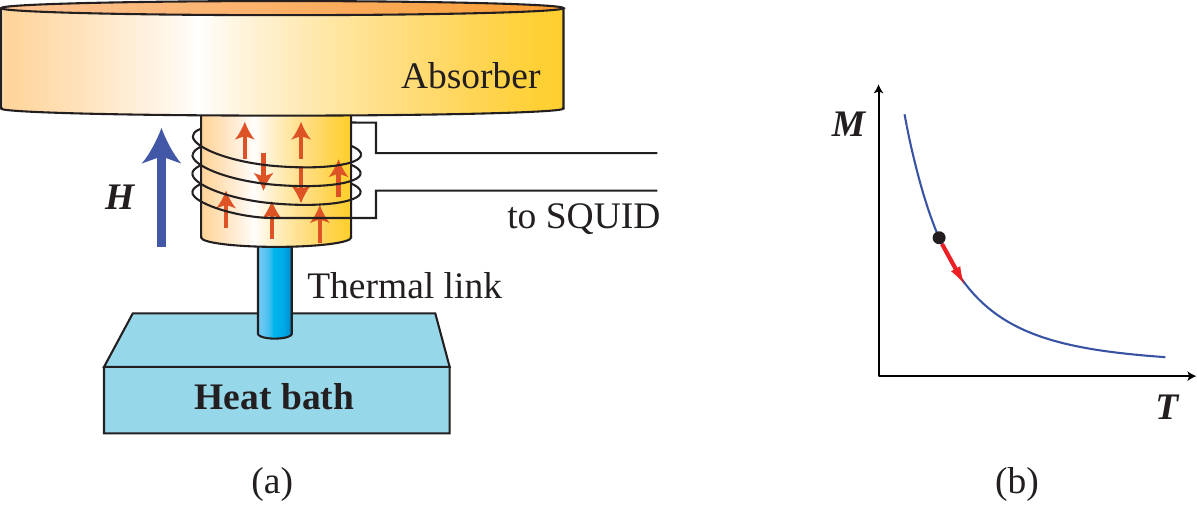}
\end{center}
\caption{(a) Simplified MMC setup with an absorber. (b) Typical $M$-$T$ characteristics of a Au:Er sensor in a magnetic field.  A temperature increase due to an input of energy  into the absorber results in a change (reduction) in the magnetization of the MMC sensor, as illustrated by the red arrow. }
\label{fig:mmc_structure}       
\end{figure}

A simplified MMC setup is illustrated in Figure~\ref{fig:mmc_structure}. Initially, MMCs for particle detection were developed in a configuration in which  a small piece of Au:Er attached to an absorber is placed inside the superconducting loop of a dc-SQUID. 
The SQUID loop itself was used as the pick-up coil of the MMC~\cite{enssbook:mmc}.
However, this early design imposed both fundamental and  practical limitations. First, it was very challenging to efficiently couple the sensor and the pick-up coil in a reproducible way. It was also almost impossible to control the thermal conductance between the sensor and the heat bath and build a detector based on carefully worked-out thermal modeling. Furthermore, the magnetic field applied to magnetize the sensor could easily affect the performance of the SQUID. 
Most of these problems were overcome by lithographically fabricating a whole MMC sensor chip. This state-of-the-art MMC design includes an integrated superconducting pick-up/magnetization coil in the vicinity of the sensor layer, maximizing the coupling between them and allowing the use of ultralow-noise SQUIDs without their performance being affected by the magnetic field used for the operation of the MMC. The thermal link between the sensor and the heat bath is now controllable by on-chip thermalization pads. 

MMCs have many advantages for use  as sensors in rare event search experiments. First, they have rather large heat capacities. Counterintuitively, this makes an MMC suitable as the sensor for an absorber with a large heat capacity because its performance will be less degraded by the added heat capacity of the absorber. Second, their working temperature range is very wide, permitting flexible optimization of the detector configuration depending on the detector performance requirements~\cite{fleischmann:2000nima,gbkim2017}. MMCs also exhibit excellent detector linearity with a small deviation that can be reliably modeled.

The fast intrinsic thermalization of MMCs is another advantage for use in particle detection. The intrinsic thermalization time is affected by the interaction between conduction electrons and magnetic  ions in the sensor material. When the absorber is strongly connected to the sensor layer or the sensor layer is used as its own absorber, an MMC can have an extremely fast signal rise-time. An MMC with a gold absorber fabricated on top of the sensor layer has demonstrated a rise-time shorter than 100\,ns~\cite{rotzinger2008,fleischmann2009aip}. 

In an application with a large crystal absorber, however, the rise-time of an MMC signal is much slower (on the order of ms) because it is primarily determined by the phonon generation mechanism and phonon dynamics in the crystal as well as the thermal connection between the absorber and sensor rather than by the intrinsic thermalization of the sensor~\cite{yhkim2004}. Nevertheless, even with the slow ms-scale overall rise-time with a scintillating crystal absorber, MMC signals are sufficiently fast to distinguish the pulse shape difference between electron- and alpha-induced events, which are considered signal  and background events, respectively, in \zerodbd{} experiments~\cite{gbkim2017}. Moreover,  a rise-time on this scale, which is approximately 1--2 orders of magnitude faster than that of thermistor-based detectors, can help reduce the unresolved pile-up (random coincidence) event rate, which can be one of the most significant background sources for slow detectors in \zerodbd{}  experiments~\cite{chernyak2012random}.

\subsubsection{Kinetic Inductance Devices}

A KID is a superconducting sensor with a very different detection mechanism from those of the detectors discussed above. It is considered a nonequilibrium detector and uses a change in kinetic inductance in a superconductor as its main signal.

\begin{figure}[t] 
\begin{center}
\includegraphics[width=8cm]{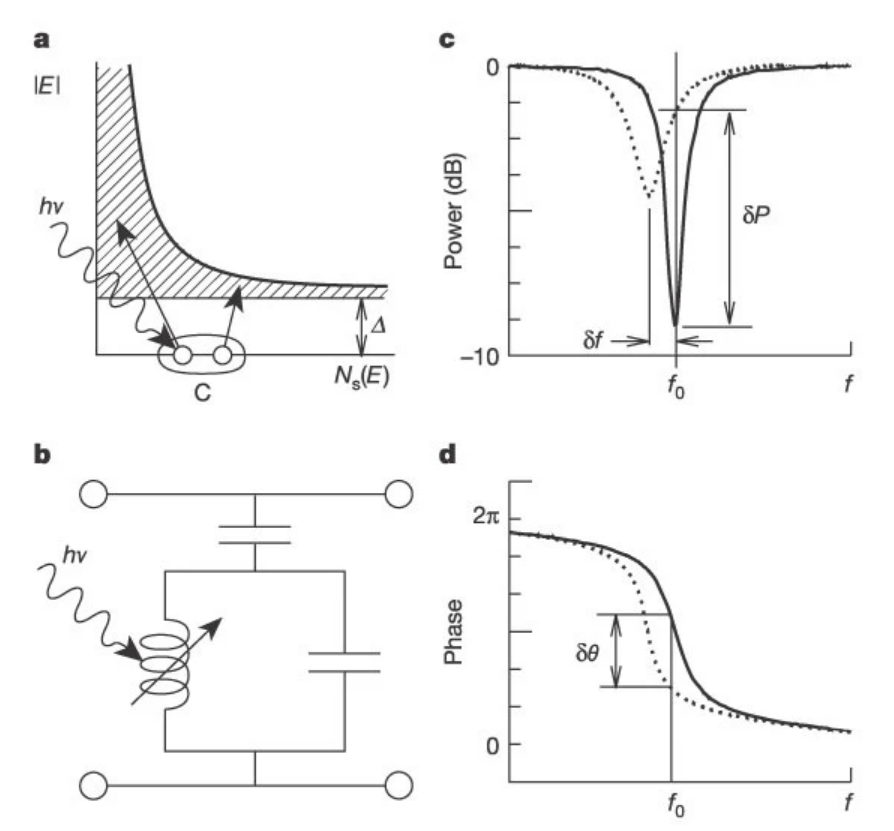}
\end{center}
\caption{Operation principle of a KID. (a) The quasiparticle density of states ($N_s$) as a function of quansiparticle energy $E$. $\Delta$ is the superconducting energy gap. A Cooper pair (denoted by C) is broken into two quasiparticles by a photon of energy $h\nu$, affecting $N_s$. (b) A schematic microwave resonant circuit of a KID. (c) Power vs.\ frequency ($f$) and (d) phase vs.\ $f$ resonance profiles of the resonant circuit at 120\,mK (solid lines) and 260\,mK (broken lines). Figure taken from \cite{day2003nature}. }
\label{fig:mkid}       
\end{figure}

Superconductors enter the resistance-free state under a dc current (smaller than a certain critical current) below a certain transition temperature. However, they exhibit a nonzero resistance under an ac current due to a complex surface impedance (or kinetic inductance) phenomenon. When a superconducting strip, cooled below its $T_{\rm c}$, absorbs energy from a particle or radiation, some of the Cooper pairs in the strip are broken, generating quasiparticles. The higher the input energy is, the more quasiparticles will be generated. These excess quasiparticles change the kinetic inductance of a microwave resonance circuit, which can be measured as a shift in the amplitude and phase of the resonance, as shown in Figure~\ref{fig:mkid}~\cite{day2003nature}. This is the basic working principle of a KID. For the superconducting strip materials in these detectors, Al, Al/Ti/Al, Hf, TiN, Ti/TiN, and PtSi are often used
~\cite{mazin2020superconducting,zobrist2019apl}.

The greatest advantage of KIDs is that they enable straightforward multiplexing of many channels. A typical KID consists of many resonance circuits with a high quality factor. By spreading the resonance frequency of each resonance circuit over a large bandwidth (e.g., 2\,MHz spacing over a 4\,GHz bandwidth), it becomes possible to send a comb of frequency tones corresponding to each resonator into the detector via a single feedline, and its response can then be measured by room-temperature electronics. In principle, more than a thousand resonators can be read out by a single feedline. Because of their highly multiplexing capability and low noise, KIDs have already been employed in many instruments to detect astronomical signals ranging from the millimeter scale to the ultraviolet (UV)~\cite{o2019energy,lee2020groundbird}.

In rare event search experiments, KIDs have not received as much attention as in astronomical applications, mainly because when an KID is integrated as a phonon detector (a thermal KID, or TKID), its energy resolution in the X-ray regime is not sufficient to be competitive with other devices such as TESs~\cite{moore2012apl}. 
However, as the focus of DM searches moves towards lowering the detection threshold, KIDs have arisen as an attractive solution. 
When a coplanar waveguide feedline with many KID resonators is fabricated on a side of an absorber wafer, the initial athermal phonons may break Cooper pairs in the resonators near the event location. Each resonator would behaves like a low- threshold phonon sensor on a target absorber helping position reconstruction of the rare events~\cite{chang2018jltp}
as in the R\&D study  of the BULky and Low-threshold Kinetic Inductance Detectors (BULLKID) project~\cite{colantoni2020bullkid}. 
Moreover, in the Cryogenic wide-Area Light Detectors with Excellent Resolution (CALDER) project~\cite{cardani2021final}, KIDs are also being developed as low-energy photon detectors for a \zerodbd{} experiment.

\subsubsection{Other technologies}

There are a few LTDs that do not use superconducting materials or superconducting circuits but are popularly used in rare event search experiments. One such detector is a doped semiconductor thermistor with a strongly temperature-dependent resistance. Neutron-transmutation-doped (NTD) Ge thermistors and ion-implanted Si thermistors are the two most common types of thermistors used at low temperatures.

When a pure semiconductor such as Si or Ge is doped at a concentration slightly below the critical doping for the metal-insulator transition, its low-temperature resistance is governed by a process called Efros-Shklovskii (ES) variable-range hopping~\cite{efros1975coulomb} and can be modeled as 
\begin{equation}
\label{eq:semi}
R(T) = R_0 \exp \left( \frac{T_0}{T} \right)^{1/2} ,
\end{equation}
with two characteristic constants, $R_0$ and $T_0$~\cite{enssbook:thermistor}. 
The resistance of a typical thermistor is 1--100\,M$\Omega$ near 50\,mK.

\begin{figure}[t] 
\begin{center}
\includegraphics[width=8cm]{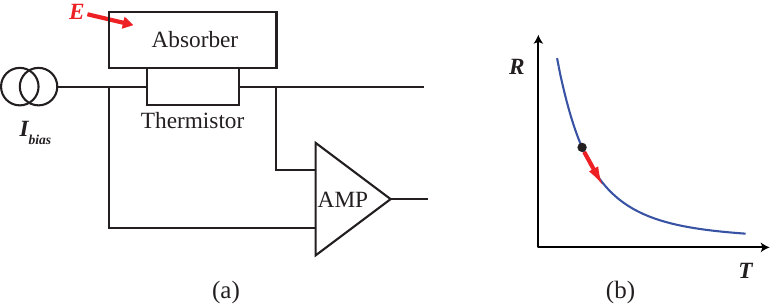}
\end{center}
\caption{(a) Simplified circuit diagram of a thermistor setup with an absorber. (b) Typical $R$-$T$ characteristics of a thermistor. An input of energy into the absorber results in a change in the resistance of the attached thermistor (represented by the red arrow). }
\label{fig:thermistor}       
\end{figure}

Such thermistors are fairly easy to use because they can be operated with conventional electronics such as junction field effect transistors (JFETs) rather than requiring sophisticated superconducting electronics. They are typically current biased,
and the measured voltage across such a thermistor is amplified by a JFET located at $\sim$100\,K or at room temperature.

Of the two types of thermistors, NTD Ge thermistors are preferred in large-scale experiments primarily because of their reproducibility and uniformity. For example, the Cryogenic Underground Observatory for Rare Events (CUORE) collaboration uses NTD Ge thermistors as the phonon sensors in their kilo-pixel ton-scale detector. On the other hand, ion-implanted Si thermistors are preferred in experiments where the individual detectors are small in size with delicate electrical and mechanical structures because the dopant ions can be implanted in a desired pattern through masking, and techniques for the micromachining of silicon are well established. Thus, Si thermistors have been a popular choice for spaceborne X-ray missions~\cite{enssbook:thermistor}.

Both Ge- and Si- based thermistors have shown reasonable energy resolutions in different energy regions. For instance, NTD Ge thermistors have achieved a FWHM resolution of 3.1\,eV at 5.9\,keV X-rays~\cite{silver2005ntd}, and Si thermistors have achieved resolutions of 3.2\,eV at 5.9\,keV and 22\,eV at 60\,keV~\cite{porter2006high} (compared to the best TES resolution of 1.3--1.6\,eV and the best MMC resolution of 1.6\,eV at 6\,keV and  10.2\,eV at 42\,keV~\cite{sikorsky2020measurement}).

One drawback of thermistors is their rather slow response time, which is caused by poor coupling between the conduction electrons and the phonons in the thermistor. This slow response is unfavorable in experiments where the rate of random coincidence events should be minimized as in \zerodbd{} search experiments.

\subsection{Heat flow model}

To maximize the performance of a cryogenic detector, it is essential to understand the heat flow mechanism between the thermal components of the detector. This is especially true for rare event search detectors with large dielectric absorbers because they involve one or more metal-dielectric, metal-metal, and/or dielectric-dielectric interfaces. Another important heat flow mechanism to be understood is the thermal coupling between the subsystems (e.g., the phonon system and the electron system) of each detector component, especially the sensor. In this subsection, three important heat flow mechanisms based on the acoustic mismatch model, electron-phonon coupling, and electronic thermal conduction will be discussed

\subsubsection{Acoustic mismatch model}
\label{subsec:sensor-amm}
The heat transfer between two different media was studied theoretically and experimentally in the 1960s--1980s~\cite{swartz1989rmp}. In particular, the heat transfer associated with thermal contact resistance, known as the Kapitza resistance (or the inverse of the Kapitza conductance), was investigated in many experiments involving liquid helium, various metals, and dielectrics.
One of the most successful models for the contact resistance was found to be the acoustic mismatch model suggested by W. A. Little \cite{little1959}. In this model, a partial transmission of phonons across an interface between two media with different densities and sound velocities is regarded as the origin of the Kapitza resistance at low temperatures. 
In other words, the thermal resistance between the two media is considered to be caused by the difference in their acoustic impedances, i.e., the acoustic mismatch.

The essential result from the acoustic mismatch model is that the thermal conductance $G_{\rm K}$ of the heat transfer is proportional to the contact area $A$ of the interface and has a cubic temperature ($T$) dependence, expressed as
\begin{equation}
G_{\rm K} = \Lambda A T^3,
\end{equation}
where $\Lambda$ is a proportionality constant that can be calculated based on  the model~\cite{swartz1989rmp,cheeke1976}. 
A number of experiments involving interfaces between various media show reasonable agreement with the predictions of this model~\cite{swartz1989rmp,pobellbook}.

\subsubsection{Electron-phonon interactions in metal films}
\label{subsec:sensor-ep}

Studies of the interactions between electrons and phonons in metals have also been a popular subject in modern physics. For instance, electron-phonon interactions play a key role in explaining the mechanism of superconductivity, which arises from indirect electron-electron interactions mediated by phonons. Another significant implication of electron-phonon interactions is that they change the effective mass of electrons in metals, offering explanations for many characteristic properties of metals. Furthermore, electron-phonon interactions are related to ultrasonic attenuation and the hot electron effect in metals.

In the context of modeling the heat flow in metals at low temperatures, we present a brief introduction to the hot electron effect originating from electron-phonon relaxation processes in metals. The hot electron effect, a temperature decoupling between the electrons and the lattice in a metal, is a phenomenon that can be observed not only at low temperatures but also at much higher temperatures. Many experiments using short laser pulses incident on metal films have demonstrated the hot electron effect. 
For example, Schoenlein et al.\ demonstrated that a high-intensity femtosecond laser pulse could induce hot electrons at a temperature higher than 1000\,K in gold, while the lattice temperature remained at approximately 300\,K~\cite{schoenlein1987femtosecond}. 
P.~B.\ Allen constructed a theory for calculating the temperature relaxation of such hot electrons via electron-phonon collisions governed by the Bloch-Boltzmann-Peierls formulas~\cite{allen1987}. The calculation assumes that the electrons first thermalize among themselves as a result of heating from the short laser pulse and then subsequently lose their energy to the lattice. An important conclusion of this
theory is that the heat transfer from the hot electrons to the lattice is proportional to ($T_{\rm e}^5-T_{\rm l}^5$), where $T_{\rm e}$ and $T_{\rm l}$ are the temperatures of the electrons and the lattice of the
metal, respectively. Allen claimed that this theory is also applicable at low temperatures.

Accordingly, the heat transfer through the electron-phonon coupling in a metal can be expressed as
\begin{equation}
\label{eq:Qep}
P_{\rm ep} = V \Sigma (T_{\rm e}^5- T_{\rm l}^5),
\end{equation}
where $V$ is the volume of the metal and  $\Sigma$ is a material dependent parameter that incorporates a coupling constant for the electron-phonon interactions in the metal. According to  Allen's theory, $\Sigma$ is approximately $2\times10^9$\,W/m$^3$K$^5$  for bulk gold based on a coupling constant found from high temperature measurements~\cite{allen1987}. 
For a small temperature difference  between $T_e$ and $T_l$,  
the corresponding thermal conductance of the electron-phonon coupling can be approximated as 
\begin{equation}
G_{\rm ep} = 5 V \Sigma T_e^4.
\end{equation}

Based on the above discussion, the heat flow between a metal film and a dielectric substrate primarily consists of two heat flow mechanisms connected in series: the Kapitza resistance from the acoustic mismatch of the materials and the impedance due to  electron-phonon interactions. The conductance arising from the series combination of these two thermal impedances is 
\begin{equation}
G=\left( \frac{1}{G_{\rm K}} + \frac{1}{G_{\rm ep}}  \right)^{-1}.
\end{equation}
Since $G_{\rm K}$ is proportional to $AT^3$, and  $G_{\rm ep}$ is proportional to $VT^4$, 
$G_{\rm K}$ governs the conductance between the metal film and the substrate at high temperatures.  However, at sufficiently low temperatures, $G$ is limited by $G_{\rm ep}$. When designing a detector with a metal film, it is often useful to know at which temperature the two conductances become equal, i.e., $G_{\rm K} \approx G_{\rm ep}$. 
For a metal film with a thickness of $t=V/A$, the equal-conductance temperature $T$ in K is approximately the inverse of $t$ in $\mu$m.

\subsubsection{Electronic heat flow}

The heat flow mechanism along a metal film or other metal structure depends on the thermal conductivity of the conduction electrons. According to diffusion theory for an electron gas, the electronic thermal conductivity $\kappa_e$ is equal to $\frac{1}{3} c v_\mathrm{F} \lambda $, where $c$ is the specific heat, $v_\mathrm{F}$ is the Fermi velocity, and  $\lambda$ is the mean free path of the electrons. At sufficiently low temperatures, because the electron mean free path is temperature independent, the thermal conductivity is proportional to the specific heat of the metal. This results in the electronic thermal conductivity being proportional to the temperature. Moreover, the electronic thermal conductivity has a simple relation to the electrical conductivity $\sigma$ for most noble metals at low temperatures, as follows:
\begin{equation}
\frac{\kappa_e} { \sigma} = L_0 T,
\end{equation}
where $L_0$ is the Lorentz number, which is equal to 2.45 $\times$ 10$^{-8}$ W$\Omega$/K$^2$. This relation is known as the Wiedemann-Franz law. It indicates that the electronic thermal conductance of a metallic structure can be estimated from  its electrical resistance and geometry. In general, the electrical resistivity $\rho$ of a metal decreases with decreasing temperature until, at very low temperatures, it reaches a finite value due to the small amounts of impurities and defects in the metal. 
The residual resistivity, $\rho_\mathrm{4K}$, also depends on the microstructure of the metal such as the grain size. When fabricating metal films for LTD applications, the residual resistivity ratio (i.e., $\rho_\mathrm{300K}/\rho_\mathrm{4K}$) is commonly used as an indicator of  film quality and is often maximized to achieve high thermal conductivity at low temperatures.

\subsection{Athermal phonon transfer processes}

In addition to the three thermal heat flow mechanisms discussed above, athermal processes also need to be well understood to build LTDs with the maximum achievable sensitivity, especially for detectors with dielectric crystal absorbers. When an energy input is deposited locally in the electron system of the absorber, initially, a cloud of energetic recoil electrons is formed, which then loses the gained energy by producing high-energy phonons. These phonons, having frequencies close to the Debye frequency of the absorber material~\cite{maris1993}, are unstable even in a perfect crystal and quickly decay into lower-energy phonons via anharmonic processes (downconversion). The lifetime of a phonon with energy $E$ is proportional to $E^{-5}$~\cite{tamura_maris1985}.

As the high-energy phonons lose their energy, their lifetime becomes longer, and accordingly, their mean free path increases. For phonons with energies corresponding to 30--50\,K in a crystal such as silicon or sapphire, the mean free path is on the order of several centimeters. Thus, for crystals of such dimensions, which are a common choice in large-scale rare event experiments, athermal phonons may travel ballistically around such a crystal until they undergo further downconversion processes from inelastic scattering off of defects or impurities (including isotope impurities) either at the surface or within the crystal bulk. This downconversion can speed up when there is either a normal metal or a superconducting feature present on the crystal surface, as is the case in most cryogenic detectors, where an LTD sensor, typically metallic, is in contact with the absorber either directly or indirectly via a thin metal or superconducting film~\cite{brandt2012jltp}. Such a film serves as a phonon collector, through which efficient energy transfer from the crystal to the sensor can occur. Thus, it is crucial to understand the downconversion decay process of  athermal phonons to optimize the heat transfer and, hence, the detector design\cite{martinez2019pra,sgkim_ltd19}.

Indeed, most of the experiments based on superconducting detectors with a crystal absorber have employed the idea of phonon collector film. This is primarily because it more efficiently transfers the energy deposited in the crystal to the sensor than when only the sensor is attached to the absorber. Furthermore, the use of phonon collectors allows one to independently optimize the dimensions of the sensor and the absorber for various absorber materials. 

\begin{figure*}[t] 
\begin{center}
\includegraphics[width=14cm]{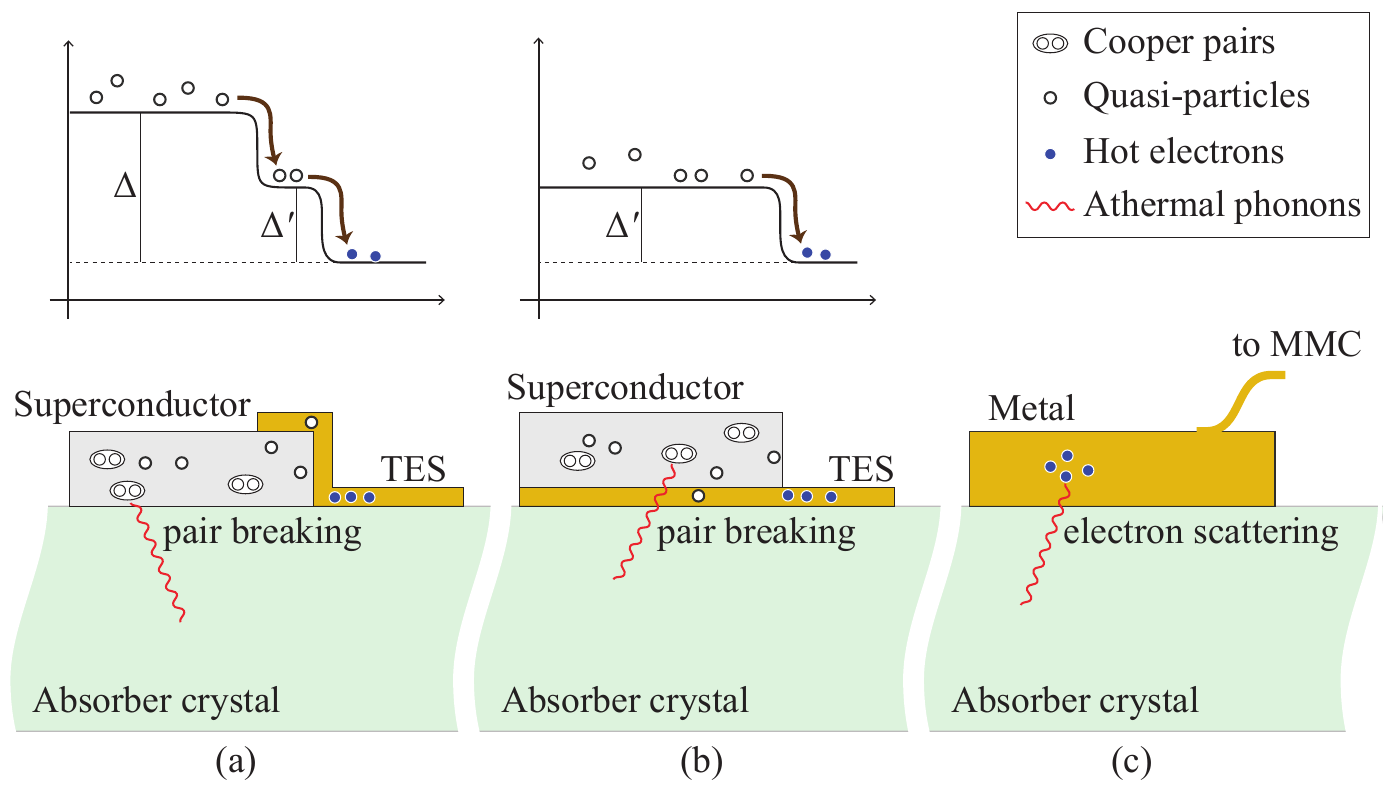}
\end{center}
\caption{Simplified energy transfer processes in a phonon collector film made of a superconductor (a and b) or a metal (c) fabricated on an absorber crystal. The energy diagrams correspond to the superconductor phonon collector films of (a) and (b). Note that $\Delta$ is the band gap of the superconductor film, and $\Delta'$ is a reduced gap originating from the proximity effect in the vicinity of a TES film operating in its transition region. A metal phonon collector such as that in (c) does not require Joule heating in the energy transfer process to an MMC.}
\label{fig:athermal_process}       
\end{figure*}

One popular choice as a phonon collector for a crystal absorber consisting of Si or \CWO{} is a superconducting Al film.
A phonon with an energy more than twice the superconducting gap of Al ($2\Delta_\mathrm{Al} \sim 340$\,$\mu$eV) may break a Cooper pair in the film, generating two quasiparticles. 
These quasiparticles diffuse along the phonon collector film to the sensor attached to the film. They then rapidly lose their energy to the electron system of the sensor, raising the temperature of the sensor. 
Another popular phonon collector is a gold film, especially when the sensor is an MMC. In this configuration, athermal phonons will directly interact with electrons in the gold film~\cite{gbkim2014}. Because of the good thermal conductance of the gold film itself and that between the sensor and the film, the energy carried by the athermal phonons is efficiently transferred to the sensor. Three different examples of phonon collectors are depicted in Figure~\ref{fig:athermal_process}.

\subsection{Phonon amplification by the Neganov-Trofimov-Luke effect}
\label{subsec:sensor-NTL}

As discussed above, the sensitivity of a detector with a heat capacity $C$ to an energy input $E_0$ is determined by the temperature increase $\Delta T =E_0/C$. Thus, for a given detector configuration, it is nearly impossible to increase the sensitivity of the detector. For a semiconductor absorber, however,  additional heat can be generated by applying an electric field, resulting in an increased energy sensitivity.
When an electric field is applied to a semiconductor, e-h pairs generated by the absorption of a particle or radiation drift along the electric field lines in the semiconductor. This drift produces so-called \textit{Luke} phonons in a process called Neganov-Trofimov-Luke (NTL) phonon amplification. In the presence of the NTL effect, the temperature increase due to an energy absorption of $E_0$ is modified to
\begin{equation}
\Delta T =  (E_0 + E_\mathrm{NTL}) / C,
\end{equation}
where $E_\mathrm{NTL}$ is the amount of  thermal energy converted from Luke phonons produced via the NTL effect. 
Since $E_\mathrm{NTL}$ is generally expected to increase with the number of created e-h pairs and the applied electric field, it can be written as
\begin{equation}
E_\mathrm{NTL} =  \eta_\mathrm{SE} \eta_\mathrm{NTL}  \cdot N \cdot e \cdot V,
\end{equation}
where $\eta_\mathrm{SE}$ and $\eta_\mathrm{NTL}$ are efficiency terms with different physical origins, 
$N$ is the number of generated e-h pairs, $e$  is the electron charge, and $V$ is the applied voltage across the electrodes. The first efficiency term, $\eta_\mathrm{SE}$, known as the suppression efficiency, depends on how well the electric field suppresses the premature recombination of e-h pairs, which depends on the gap distance and the voltage $V$ between the electrodes. The second term, $\eta_\mathrm{NTL}$, is the efficiency associated with the creation of Luke phonons. 

In the ideal case of perfect efficiencies  $\eta_\mathrm{SE}$ and $\eta_\mathrm{NTL}$, all the created e-h pairs are collected in each electrode, generating a heat dissipation equal to 
 $(e \cdot V)$ per pair. 
In reality, however, the created e-h pairs may be recombined or trapped by impurities inside the semiconductor or due to imperfect surface conditions, causing a reduction in the generated heat. 
Thus, the amplification gain of the NTL effect can be expressed as
\begin{equation}
G_\mathrm{NTL} =  \Delta T / \Delta T_{(V=0)}  = 1 + \eta \frac {eV}{\epsilon},
\label{eq:NL_gain}
\end{equation}
where  
$\eta$ is the product of $\eta_\mathrm{SE}$ and  $\eta_\mathrm{NTL}$ and
$\epsilon$ is the average energy needed to generate an e-h pair.  With a sufficiently large $V$, the gain becomes proportional to $V$. However, at much higher voltages, the gain will saturate because the combined efficiency also depends on $V$ and, in general, degrades at high voltages~\cite{isaila2012low}. Since the detector noise also tends to increase with increasing voltage, an optimal voltage should be found to maximally utilize the NTL effect.

The ability to increase the detector sensitivity by means of the NTL effect has been demonstrated in a number of applications based on various LTD sensors, including TESs, MMCs and NTD Ge thermistors~\cite{defay2016jltp,cuore_luke,romani2018apl,jajeon2020jltp}. These detectors showed reasonable amplification gains without significant degradation in noise level, resulting in improved sensitivity. In particular, sensitivity to single e-h pairs was recently demonstrated for visible photons with an applied voltage of up to 160\,V in a silicon crystal~\cite{romani2018apl}. This achievement may have a large impact, particularly in low-mass DM search experiments.

In early-generation DM detectors based on calorimetric measurements with Si or Ge absorbers, charge signals were measured together with heat signals~\cite{cdms2,edelweiss2}. The primary reason for the simultaneous detection of charge and heat was event discrimination between electron recoils and nuclear recoils. Since those early-generation experiments were looking for heavy DM particles that would interact more strongly with nucleons than with electrons, they used a small bias voltage that could efficiently reject electron recoils. However, as the focus of these experiments has shifted towards the detection of
 low-mass DM particles, the main goal of detector design has become the lowering of the energy threshold. This can be achieved by utilizing the NTL effect with a much higher bias voltage, although as a trade-off, the detector becomes incapable of event discrimination. Using single-charge-sensitive silicon detectors, SuperCDMS has recently published significantly improved DM constraints in the low-mass region~\cite{agnese2018prl}.

 
\begin{figure*}[t] 
\begin{center}
\includegraphics[width=16cm]{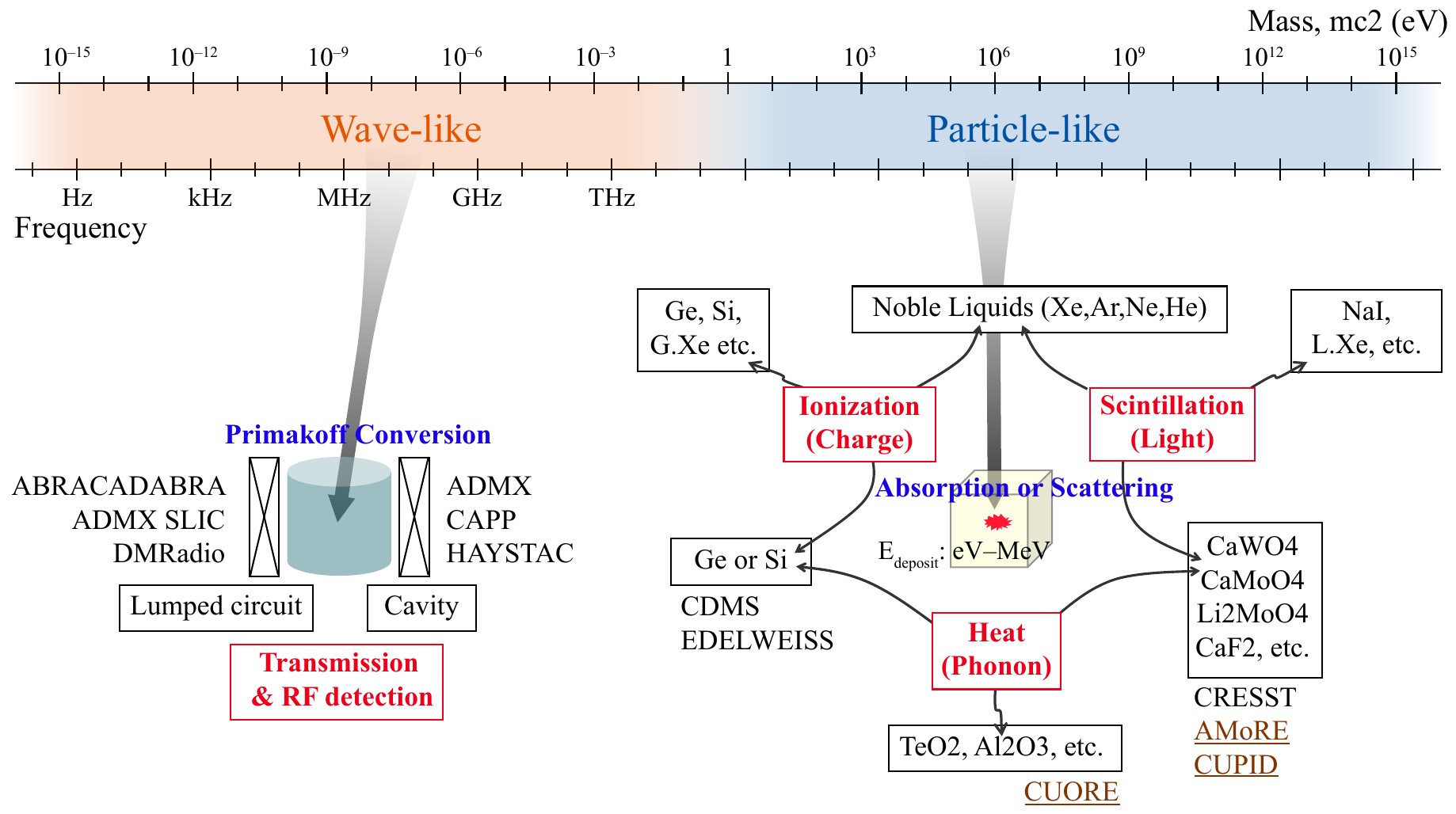}
\end{center}
\caption{Mass range of DM (\textit{top}) and various detection methods with selected ongoing projects (\textit{bottom}). Particle-like DM is expected to deposit eV--MeV-scale energy in the detector, resulting in ionization, scintillation, and/or heat signals, as shown in the red boxes. While some experiments utilize a single type of signal, others simultaneously measure two types of signals to improve the detection sensitivity. Commonly used absorber materials are listed in the black boxes for six different detection schemes. Ongoing DM search projects and \zerodbd{} projects (brown text) that utilize heat signals are also listed next to the corresponding absorber materials. At the bottom left, detection methods that utilize low-temperature microwave cavities and superconducting lumped circuits are listed along with ongoing wave-like DM search projects.} 
\label{fig:detection_methods}       
\end{figure*}

\section{Application of superconducting detectors for rare event searches}
\label{sec:app}

In this section, we review how the superconducting sensor technologies discussed in Section~\ref{sec:sensor} can be used to develop a highly sensitive superconducting detector for rare event search experiments by discussing several real-life examples.

As discussed in Section~\ref{sec:challenges}, despite the abundance of compelling but indirect evidence for the existence of DM, no direct detection has been achieved to provide insight into the individual mass and coupling strength of DM. Thus, an overwhelming number of theories have been proposed to explain the observed indirect evidence and predict the properties of the DM. This situation has resulted in a vast range of allowed masses for DM that span many orders of magnitude, as shown in Figure~\ref{fig:dm_candidate} and in Figure~\ref{fig:detection_methods}. Since there is no single detection method that can cover such a wide range, a detector should be optimized for a rather narrow mass range. In Figure~\ref{fig:detection_methods}, we provide a list of detection methods suitable for the direct detection of DM of different mass ranges as well as for \zerodbd{} search experiments that share common detection schemes with the direct DM search experiments. 

First, the detection of DM in the mass range above 1\,eV relies on the \textit{particle}-like nature of DM. Such DM is expected to deposit energy in the range of eV--MeV into the detector as a result of interactions between the DM and the absorber material. 
Depending on the type of absorber material, this energy deposition can result in ionization (charge), scintillation (light) and/or heat (phonon) signals. 
The bottom right of Figure~\ref{fig:detection_methods} shows six different categories of target materials that allow single- or dual-channel measurements. Dual-channel measurements have the advantage of enhanced background rejection via particle identification, as explained below.

Superconducting detectors that utilize the heat (phonon) channel as their main signal have excellent promise for use in particle-like DM searches because such detectors have demonstrated superior energy resolution and a lower detection threshold in the eV--MeV energy range. Moreover, the ubiquity of phonon signals facilitates the development of a detector with an absorber that allows dual-channel measurements. 

\begin{figure}[t] 
\begin{center}
\includegraphics[width=6cm]{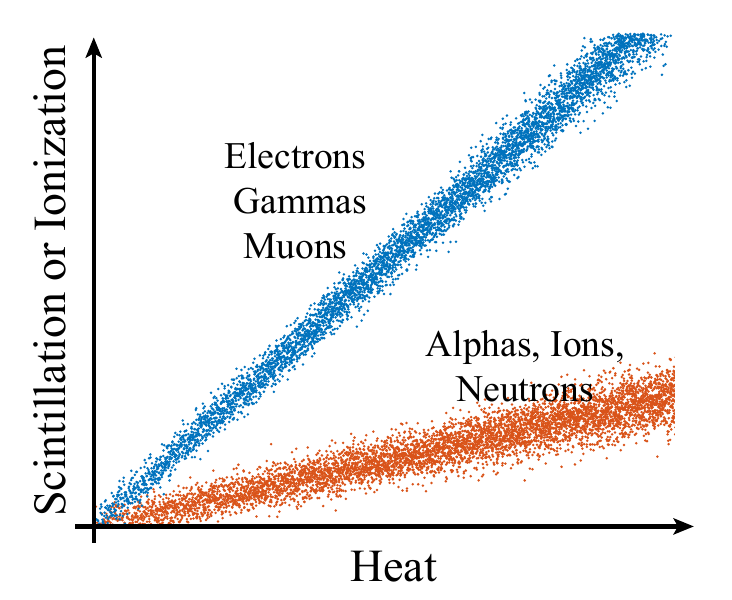}
\end{center}
\caption{ The concept of particle identification in the dual-channel detection of heat-scintillation or heat-ionization. 
Different types of particles depositing energy in a target absorber can result in different ratios of scintillation or ionization to heat signal amplitudes. For simplicity, the events are grouped in two, one resulting from electrons, gammas, or muons (blue) and the other from alphas, ions, or neutrons (red). Depending on the absorber material and detection technique, the latter group can be further separated. 
}
\label{fig:particle_id}       
\end{figure}

In dual-channel detection methods (i.e., heat-ionization, heat-scintillation, and ionization-scintillation), the ratio of the amplitudes of the two channels differs depending upon the type of particle that primarily receives energy upon interaction with a DM particle. For instance, in the heat-scintillation case, an energetic ion that results from nuclear recoil due to DM-nucleon scattering generates much less scintillation light than electrons or photons (e/$\gamma$) of the same energy, as illustrated by Figure~\ref{fig:particle_id}. This effect is known as quenching of a scintillating material~\cite{tretyak2010semi}. Similarly, in a semiconductor, energetic ions generate fewer e-h pairs than electrons with the same energy because they also lose their energy to phonons or other nuclei. 

Dual-channel measurements have also been used in \zerodbd{} search experiments, where the main signal arises from two electrons with energy of a few MeV produced by the decay. In these experiments, alpha-emitting radionuclides inside and near the absorber crystals are crucial background sources that can be efficiently removed by dual-channel detection.

The detection of DM candidates with masses below 1\,eV is mostly based on the wave-like nature of the DM.
For instance, a tunable resonator composed of a microwave cavity or a superconducting lumped circuit has been developed to detect QCD axions, a strong DM candidate, through their conversion to photons under a high magnetic field (Primakoff conversion)~\cite{1984:Sikivie}. This type of detection method, along with several ongoing projects, is illustrated at the bottom left of Figure~\ref{fig:detection_methods}. Note that other promising technologies exist~\cite{battaglieri2017} that are not listed here, some of which we briefly discuss later in the section. 




\subsection{DM detectors} 
\label{subsec:app-dm}

	\subsubsection{CDMS}  
The Cryogenic Dark Matter Search (CDMS) is an international collaboration for direct detection of WIMPs that has pioneered many important superconducting detector technologies. In particular, it is well known for its major contributions to the development of TES and heat-ionization dual-channel detectors based on quasiparticle diffusion in thin films and the NTL effect in semiconductor absorbers. Most of these techniques were already discussed in the previous section, but we will briefly revisit them when applicable.

The first phase of CDMS was called CDMS I (1998-2002), which was followed by CDMS II (2003--2009) and then  SuperCDMS Soudan (2011--2015). The current phase is SuperCDMS SNOLAB, and a detector is being built in the deep underground laboratory at SNOLAB. 

With each phase of the experiment, the number and the total mass of the target crystals increased. CDMS I included 6 detector modules composed of a Ge crystal, with a total mass of 1\,kg. The experiment was conducted at a shallow underground laboratory (SUF, $\sim10$ mwe (meter water equivalent)) and achieved a total exposure of 30\,kgd. CDMS II included 30 detector modules (4\,kg of Ge in total) in a deeper underground laboratory of $\sim2100$\,mwe in Soudan. With 400 kgd of exposure,  CDMS II resulted in the then strongest limit on the WIMP-nucleon spin-independent scattering cross section, nearly rejecting the DAMA/LIBRA annual modulation region. 
SuperCDMS Soudan included 15 detector modules with a total Ge mass of 9 kg and was also conducted in Soudan. The next-generation experiment SuperCDMS SNOLAB will be operated on the order of 100 detector modules totaling approximately 200\,kg. This latest phase aims to probe not only the traditional WIMP mass region but also the light DM region. For the latter, a fraction of the detectors are configured to have an ultralow detection threshold.

With each successive phase, the CDMS detector design also continued to evolve. However, the core measurement scheme remained the same: the detector consists of multiple detector modules operating at cryogenic temperatures, each module is equipped with a sizeable disk-shaped Ge or Si crystal as the target for DM-normal matter interactions, and the temperature and ionization signals induced by an energy input in the target crystal are simultaneously measured by a temperature sensor and a field effect transistor (FET) amplifier, respectively, as illustrated in Figure~\ref{fig:detection_methods}. The dual-channel measurement makes it possible to distinguish nuclear recoil (NR) signals from electron recoil signals, thus significantly improving the detector sensitivity to NR-like DM signals, as previously discussed (see Figure~\ref{fig:particle_id}).

\begin{figure}
\begin{center}
\includegraphics[width=1.0\columnwidth]{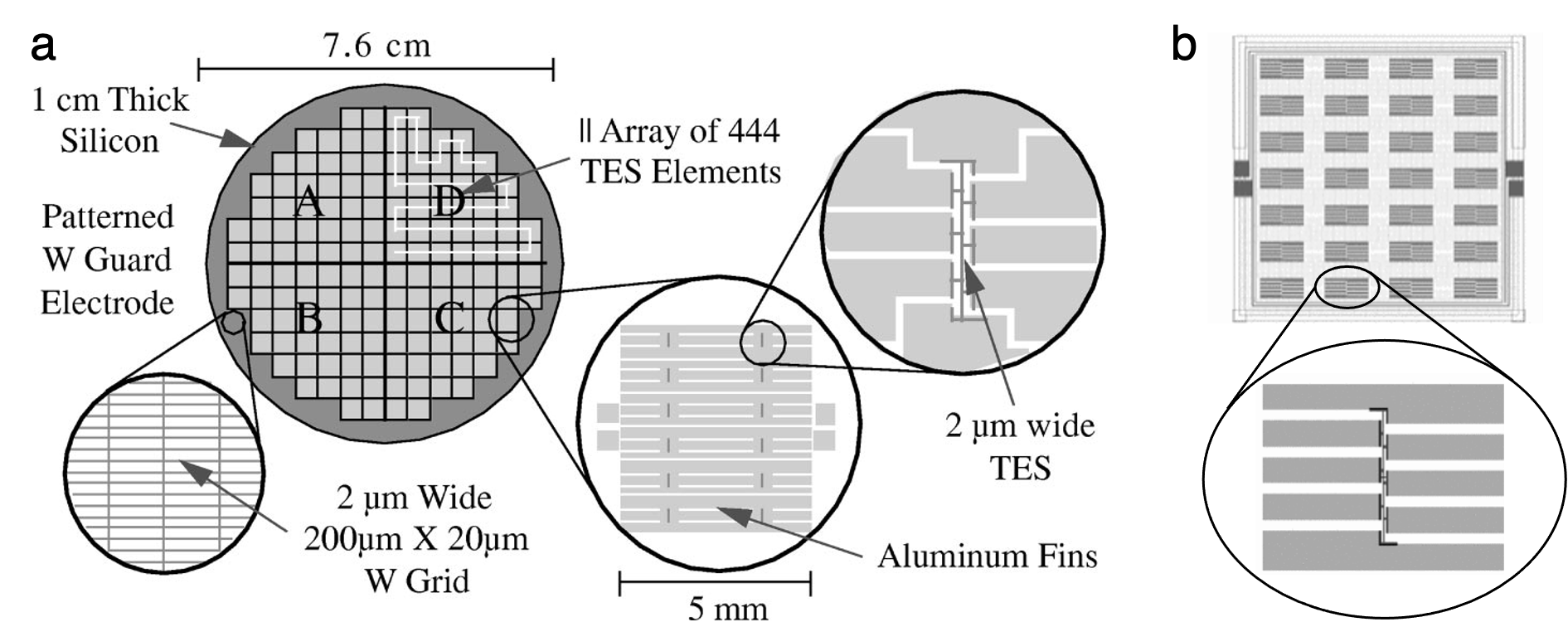}
\end{center}
\caption{(a) Diagram of the phonon side of the CDMS I ZIP detector with a 100 g Si target crystal. Diagram of the phonon sensors for the 100\,g Si ZIP
detector. The phonon sensor is divided into 4 channels labeled A to D. Each channel consists of 37 cells, and each 5 mm $\times$ 5 mm cell consists of 12 TES elements connected in parallel. The TESs are coupled to Al quasiparticle-collector fins. The figure is taken from Ref.~\cite{Abrams:2002nb}. (b) Diagram of a cell consisting of 28 TES elements in the CDMS II ZIP detector. The figure is taken and modified from Ref.~\cite{akerib2005exclusion}.
}
\label{fig:phonon_sensor_zip}       
\end{figure}

The temperature sensors in the CDMS detectors, except for that in one early generation called the Berkeley Large Ionization- and Phonon-mediated (BLIP) detector, which is based on NTD Ge sensors, consist of thousands of tiny W TESs spread over the surface of the target crystal in a hierarchical arrangement, a unique feature of CDMS's Z-sensitive ionization and phonon-mediated (ZIP) detector. More specifically, CDMS I ZIP contained 4 independent phonon channels, with each channel consisting of 37 cells and each cell consisting of 12 TES elements coupled to Al ``fins'' or 12 quasiparticle-assisted electrothermal feedback transition-edge sensors (QETs) as shown in Figure~\ref{fig:phonon_sensor_zip} (a).%
In total, there were $4\times444$ TES elements for each target crystal. The TESs had a $T_{\rm c}$ of $\sim100$\,mK and were voltage-biased such that they were in the extreme electrothermal feedback mode. This ensured stable operation of such a large number of TES elements.
The Al fins served as quasiparticle collectors as well as the ground electrode of the ionization detector and covered 82\% of the top surface of the crystal. The phonon signal was read out by SQUID arrays.

The CDMS II ZIP detector had a layout very similar to that of the CDMS I ZIP detector but with an improved phonon collection efficiency. The QETs of the CDMS I ZIP detector had a fill factor of nearly 100\% near the QET coverage area (82\% of the top surface); because the average distance between generated quasiparticles and the nearest TES was larger than the diffusion length of the quasiparticles, some quasiparticles were unable to reach a TES within the time constant of the TES. To solve this problem, the number of TESs per channel was increased from 12 to 28, and the length of the Al fins was decreased in the CDMS II ZIP detector as shown in Figure~\ref{fig:phonon_sensor_zip} (b). There were even more upgrades in the SuperCDMS detectors, as described below.

The ionization channel of the CDMS detectors consists of charge-collection electrodes deposited on both sides of the target crystal, which are used to generate the electric field needed for the created e-h pairs to drift along the field lines and to be collected in the electrodes~\cite{Shutt:2000gg}.
To reduce the loss of the ionization signal for surface events, a thin layer of amorphous silicon was deposited between the electrode and the target crystal in every detector except CDMS I ZIP~\cite{Abrams:2002nb}.  In the CDMS I and II detectors, the electrode on the top surface (the ``phonon side'') served as the ground electrode, and the electrode on the bottom surface (the ``ionization side'') was biased and connected to a current integrator. The generated e-h pairs under electric fields created NTL phonons, which contributed to the total phonon signal observed by the phonon sensor, as illustrated in Sec.~\ref{subsec:sensor-NTL}. To prevent a gradual drop in gain of the NTL phonon amplification due to charged trap centers in the crystals, LEDs were used to routinely neutralize the crystals. Moreover, to avoid creating too many NTL phonons, which would compromise the detector's PID capability, a relatively small bias voltage of a few V was used in CDMS I and II.

In addition to exhibiting PID capability, the CDMS detectors have position sensitivity, which can be utilized to further reject the background signal. As described above, the CDMS I and II ZIP detectors have phonon sensors segmented into four individual channels. By analyzing the pulse shape of a phonon signal from the four channels (e.g., by comparing the arrival times), the lateral position of the signal can be determined. Furthermore, the pulse-shape difference between the bulk and surface events can be used to reject the background signal occurring in the detector's ``dead layer'' near the surface, where the sensitivity of the ionization channel is compromised.

\begin{figure}
\begin{center}
\includegraphics[width=1.0\columnwidth]{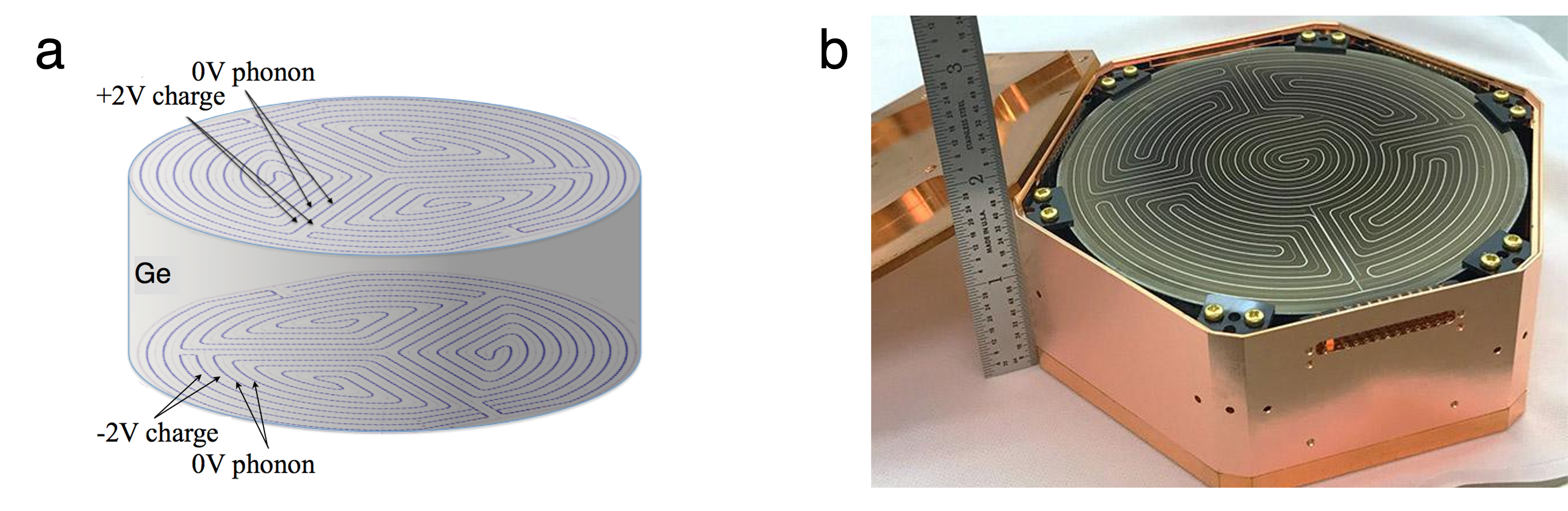}
\end{center}
\caption{(a) Diagram of an iZIP detector with a crystal 76\,mm in diameter and 25\,mm thick used in  SuperCDMS Soudan. 
(b) Photograph of an iZIP detector with a crystal 100\,mm in diameter and 33\,mm thick and a copper housing to be used in SuperCDMS SNOLAB. The figure was taken and modified from Ref.~\cite{Agnese:2013ixa}.
}
\label{fig:izip}       
\end{figure}

Now, we describe the SuperCDMS detectors. Based on the basic working principle of the CDMS I and II ZIP detectors, SuperCDMS has involved significant upgrades in its new interleaved Z-sensitive Ionization and Phonon sensors (iZIP) detector. In this new detector, phonon sensors are patterned on both sides of the crystal and interleaved with ionization electrode lines, as shown in Figure~\ref{fig:izip}.%
The phonon sensors on the top and bottom surfaces are separated into 4 channels in the SuperCDMS Soudan iZIP and 6 channels in the SuperCDMS SNOLAB iZIP, covering the entire top and bottom surfaces of the crystal. The phonon sensors also serve as the ground electrodes in both ZIPs and iZIPs. However, unlike ZIP detectors, iZIP detectors have both positive (+2\,V) and negative ($-2$\,V) bias electrodes. These changes were made to improve the sensitivity of the iZIP detectors to surface events and to better extract the energy and position information of each event. 

There are several variants of the SuperCDMS Soudan and SNOLAB iZIP detectors, such as CDMSlite and CDMS-HV. The main characteristic of these variants is that they use a much higher bias voltage ($> 50$\,V) rather than a few V. This higher voltage increases the gain of NTL amplification, thus improving the energy resolution of the phonon channel and decreasing the energy threshold at the expense of compromised PID capability. This was intended to enable probing of the low mass region of the DM parameter space. SuperCDMS-CPD, a new detector developed in collaboration with the Cryogenic PhotoDetector  (CPD) project~\cite{fink2021performance}, has taken a different approach. SuperCDMS-CPD uses a small (approximately 10\,g) Si crystal rather than massive Ge crystals. The use of such a small crystal resulted in a substantially improved baseline energy resolution of approximately 3.9\,eV~\cite{Alkhatib:2020slm}, which resulted in a promising limit curve, as shown in Figure~\ref{fig:dm_limit}.

	\subsubsection{CRESST}	


The Cryogenic Rare Event Search with Superconducting Thermometers (CRESST) is another multi-institution direct DM search experiment that uses TESs as sensors of the DM-normal matter interaction. CRESST is known for its use of scintillating crystals equipped with phonon-photon channel detection, as shown in Figure~\ref{fig:detection_methods}. The collaboration recently completed the operation of CRESST-III Phase 1~\cite{petricca2020first} and is in the preparation stage for CRESST-III Phase 2. 

\begin{figure}[b] 
\floatbox[{\capbeside\thisfloatsetup{capbesideposition={right,top},capbesidewidth=4.5cm}}]{figure}[\FBwidth]
{\caption{Schematic diagram of the active thermal feedback scheme. The Au wire was used to make thermal contact between the heater and the tungsten thermometer. The substrate was an Al$_2$O$_3$ crystal. The figure was taken from Ref.~\cite{altmann2001results}}\label{fig:cresst1_layout}}
{\includegraphics[width=3cm]{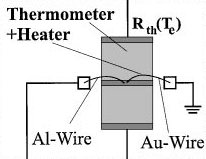}}
\end{figure}

The first CRESST experiment (CRESST-I) used four sapphire (Al$_2$O$_3$) crystals (262\,g each) as the target material for the detection of DM-nucleon scattering events~\cite{ferger1994massive}. Sapphire crystals were chosen because of their high Debye temperature~\cite{berman1955thermal}, which can make the specific heat of the detector very small at cryogenic temperatures. To increase the resolution and sensitivity of the detector, the operating temperature was decreased to as low as $\sim15$\,mK. To match this low temperature, a TES composed of a W film (or a superconducting phase transition (SPT) thermometer, as it was called within the collaboration~\cite{seidel1990phase}) was developed. To achieve this very low $T_\mathrm{c}$, an $\alpha$-phase W film was electron beam evaporated on heated sapphire crystals in an ultrahigh vacuum environment~\cite{colling1995low}. For stable operation of the detector at the $T_\mathrm{c}$ of the W film of $\sim15$\,mK, an active thermal feedback scheme in which the base temperature of the cryostat was maintained at $\sim6.5$\,mK and an additional heater integrated into the detector maintained the detector temperature at the $T_\mathrm{c}$~\cite{meier2000active} was adopted, as shown in Figure~\ref{fig:cresst1_layout}.%
The heater also served as a means to calibrate the detector.

The first DM limit result from the CRESST-I detector was released in 2002~\cite{angloher2002limits}. Although these detectors had the highest sensitivity per unit mass among cryogenic detectors in use at the time, this result did not have sufficient sensitivity to rule out the ``DAMA'' region of WIMP's annual modulation. This limitation was mainly ascribed to the inability to discriminate NR-like signals from gamma-dominated background events. The collaboration implemented a PID solution to increase the detection sensitivity by employing simultaneous detection of heat (phonon) and light (photon) signals~\cite{meunier1999discrimination}, which is a slightly different scheme from that of the CDMS, whose PID mechanism is based on simultaneous detection of heat (phonon) and ionization (charge) signals.

\begin{figure*} 
\includegraphics[width=11cm]{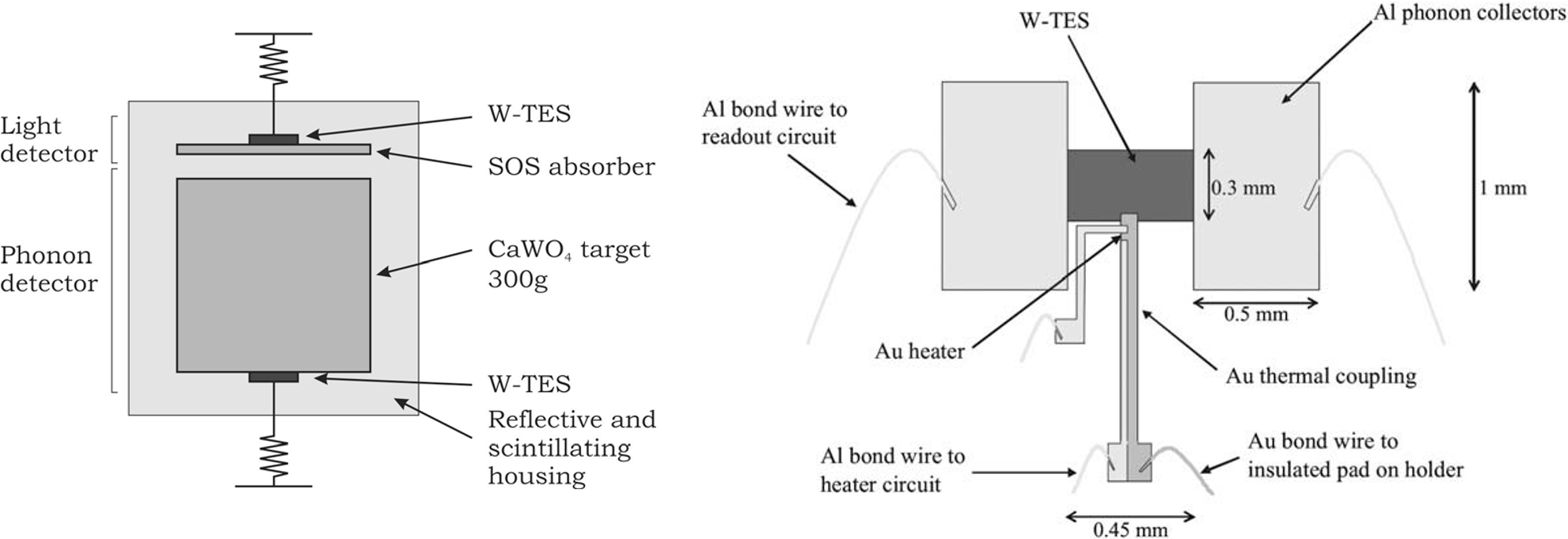}
\caption{\label{fig:cresst2_setup}
({\it left}) Heat and light detector module. ({\it right}) Thermometer-heater layout for the light detector. The figure was adopted from Ref.~\cite{angloher2009commissioning}.
}
\end{figure*}

In certain scintillating crystals, electron- and photon-induced signals show a different light yield than that due to heavy ions or neutrons, as shown in Figure~\ref{fig:particle_id}.
Among several different scintillating crystals the CRESST collaboration tested, CaWO$_4$ showed the best performance at cryogenic temperatures. 
Based on this finding, they made significant changes in their detector design (CRESST-II): the sapphire crystals were replaced with CaWO$_4$, and light detectors were added, as shown in Figure~\ref{fig:cresst2_setup}.%

Note that the layout of the light detector is slightly different from that of the phonon detector. As shown on the right side of the figure,  Al phonon collectors with large areas, which were meant to maximize the signal size for a given energy deposition, were added.

The results of the commissioning phase of CRESST-II (completed in 2007) as well as those of the subsequent CRESST-II phase 1 (2009--2011) showed improved DM limits that excluded the DAMA region~\cite{angloher2009commissioning,brown2012extending,angloher2012results}. 
This result was consistent with the null finding of other DM detectors, including the two LTD experiments of CDMS and EDELWEISS. 
The design of the detector module was updated in four different ways in CRESST-II phase 2 to clarify some events with no or small amplitudes in the light channels that were unlikely from DM scatterings or leakage events due to known backgrounds. One 300\,g module of CRESST-II phase 2 showed a 300\,eV threshold, resulting in a large improvement in the DM detection sensitivity~\cite{angloher2016results, angloher2019limits}.
However, CRESST-II observed excess events in the ``acceptance'' region in both experimental phases, which were interpreted as background instead of  positive WIMP signals.   

\begin{figure}[b] 
\includegraphics[width=6cm]{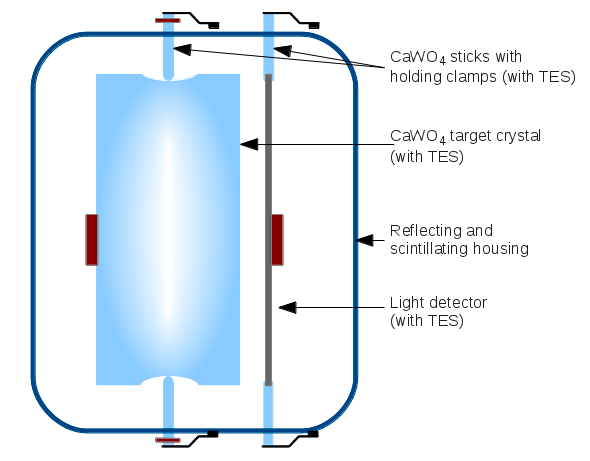}
\caption{\label{fig:cresst3}
Schematic diagram of the CRESST-III detector module. Note that the detector holders are equipped with additional TESs as veto detectors to reject events associated with the crystal holders. The figure was taken from Ref.~\cite{petricca2020first}.
}
\end{figure}

The most recently completed CRESST-III phase 1 (2016--2018) focused on the low-mass WIMP region. For this purpose, the mass of the CaWO$_4$ crystals was reduced to 24\,g each (a factor of 10 reduction), significantly lowering the detector threshold to below 50\,eV~\cite{abdelhameed2019prd}. 
Moreover, the new modules were designed to employ crystal supporting rods composed of CaWO$_4$, and each rod had a TES film, as shown in Figure~\ref{fig:cresst3}. This was to veto any events caused by the thermal signal propagating to the target crystal through its holding structure, which could explain the excess events in the acceptance region. 

CRESST-III phase 1, with its lower threshold, resulted in an improvement in the DM limit in the low-mass region ($\sim500$\,MeV/c$^2$ and below) by an order of magnitude. A new run of CRESST-III phase 2 with considerably increased exposure is planned, which is expected to further extend the low-mass region of the DM parameter space.

Based on the sensor technology developed for CRESST, another DM search project, COSINUS, has been proposed to cross validate the 14-year long annual modulation results of the DAMA/LIBRA~\cite{angloher2020jltp,angloher2021simulation}. This project is aiming to develop a heat-light detector setup with a detection threshold as low as 1\,keV using CRESST-style TES sensors and NaI scintillating crystals, a type of crystals that DAMA/Libra measured with conventional photo-multiplier tubes~\cite{bernabei2013epjc}.

	\subsubsection{Other low-temperature DM detectors}	

Using TESs, CDMS and CRESST have shown remarkable progress and promising results in direct DM detection covering the low-mass region of the DM parameter space.  Exp\'erience pour DEtecter Les WIMPs En Site Souterrain (EDELWEISS), another DM search project based on LTDs, utilizes NTD Ge thermistors coupled to Ge target crystals as the phonon sensor. The Ge crystals are also equipped with electrodes to measure the ionization signal. This two-channel detection allows distinction of nuclear recoil events from electron recoils in the target crystal. 

EDELWEISS included three measurement phases, all of which were carried out at the Modane Underground Laboratory in France. 
EDELWEISS-I, the first phase, consisted of three detector modules composed of 320\,g Ge crystals (approximately 1\,kg in total). EDELWEISS-I resulted in the first rejection of DAMA's positive claim of DM detection~\cite{sanglard2005final}. After this result, many other experiments, including CDMS and CRESST, also ruled out DAMA's claim. In EDELWEISS-II, the collaboration operated 10 detector modules of 400\,g Ge crystals with a sensitivity of $4.4\times10^{-32}$\,cm$^2$ for a DM mass of 85\,GeV~\cite{armengaud2011final}. In EDELWEISS-III, 24 detector modules of 820--890\,g Ge crystals were operated to search for DM in the mass range of 1--20\,GeV~\cite{gascon2020low}. All three phases yielded null results of the DM signal. Currently, as with CDMS and CRESST, the EDELWEISS collaboration is shifting its focus to the detection of low-mass DM~\cite{gascon2020low}. 

EDELWEISS is acclaimed for the development of an efficient method to reject surface events, which are one of major limiting factors of the sensitivity of DM detection  in the phonon channel. Surface events on a Si or Ge target crystal may undergo incomplete charge collection, which can result in misidentification of electron-induced surface events by nuclear recoil events because, although due to a different mechanism, nuclear recoils also give rise to a relatively small ionization signal for a given energy input. To solve this issue, EDELWEISS designed a new detector with two sets of interdigitated electrodes patterned on the surface of the Ge crystal~\cite{defay2008cryogenic,broniatowski2008cryogenic}. One set of electrodes is used to collect ionization signals from bulk events, and the other set is used to veto surface events. In the latest version of the detector, called the fully interdigitated detector (FID), the electrodes cover every surface of a crystal to improve the event discrimination efficiency. The FID has demonstrated more than 99\% efficiency~\cite{Armengaud:2017nkf}. Similar interdigitating electrodes were also adopted in SuperCDMS's iZIP detectors, as discussed in the previous section.

	\subsubsection{Nanometer-scale thermal calorimeters}

During recent decades, significant progress has been made in the development and understanding of low-dimensional devices~\cite{smith1996low}. Many novel low-dimensional devices are based on superconducting technologies and operate at sub-Kelvin temperatures. 
These devices were motivated by various practical applications, such as quantum computing and memory devices, but have also found their way into astroparticle physics as detectors of rare event search experiments. 
Here, we briefly introduce detectors based on low-dimensional devices (low-D detectors).

In thermal calorimetric detection, a smaller heat capacity increases the signal size, which is set by the temperature change for a given input energy, as described in Eq.~\ref{eq:typical}. One way to maximize the temperature increase is to make a detector without an absorber and use the sensor itself as an absorber. The heat capacity of such detectors can be further reduced by shrinking the dimensions of the sensors. Thus, low-D nanometer-scale detectors have become a promising choice for calorimetric detection of DM.

These nanometer-scale thermal detectors, however, are not always a viable solution for the direct detection of DM, especially for WIMP-like DM particles with a mass of approximately 100\,GeV, because the small overall size of the detector prevents it from being competitive in this mass region. On the other hand, they can be an excellent choice for light DM because of their extremely low detection threshold. Low-D detectors based on tunnel junctions are good examples. 

The Josephson-threshold calorimeter (JTC) is a low-temperature thermal detector based on a temperature-biased tunnel Josephson junction formed by different superconductors~\cite{guarcello2019pra}.
A Josephson junction is a device consisting of two or more superconductors connected by a weak link typically made of an insulator or a normal metal that operates based on the Josephson effect~\cite{josephson1962possible}.
Josephson junctions have tremendous applications, especially in quantum sensing.
For JTCs, Josephson junctions engineered with two superconductors of different $T_{\rm c}$'s  and different superconducting gap $\Delta$'s are employed to have a sharp critical current dependence  on the temperature. 
Using the lower $T_{\rm c}$ junction element as the sensor/absorber, JTCs are expected to be capable of photon number resolution for photons from the mid-infrared region through ultraviolet region (frequencies in the range of 30\,THz to $9\times10^4$\,THz). 

The graphene Josephson junction (GJJ) detector is another type of low-D detector suitable for light DM. A GJJ uses a graphene film as the sensing element~\cite{ghlee2020nature,walsh2021science}. It utilizes the peculiar property of graphene in which the electron density of states vanishes at the charge-neutrality point, which leads to a highly suppressed electronic specific heat~\cite{geim2009science}. 
Recently, a DM experiment was proposed based on GJJ detectors that can probe the DM parameter space down to 0.1\,keV~\cite{djkim2020arxiv}.

The quantum capacitance detector is another low-temperature detector capable of low-energy photon counting. When a photon is absorbed in a mesh absorber composed of superconducting aluminum, the photon produces free electrons, which tunnel into a single Cooper-pair box (SCB) island, a variable capacitor embedded in a resonant circuit. A change in the SCB state due to electron tunneling is recorded an RF signal. Recently,  THz single photon detection was demonstrated using this technique~\cite{echternach2018na}. 

These small low-temperature devices are mostly in development and have not yet been deployed in an active DM search experiment. However, as an increasing number of DM search experiments report null results in the traditional WIMP-like DM mass region and the focus shifts toward lower mass DM, these detectors are expected to play a crucial role in next-generation DM search experiments.

\subsection{Application to neutrinoless double beta decay search} 
\label{subsec:app-dbd}


As discussed in Section~\ref{subsec:dbd_intro}, the discovery of the \zerodbd{} process  answers several fundamental questions in particle physics and astrophysics. Because of the importance of this process,  \zerodbd{} searches have become  one of the most sought-after  rare event experiments conducted in deep underground laboratories. 
Among many detection technologies, thermal calorimeters have played a major role in \zerodbd{} searches during the last two decades~\cite{avignone2008rmp,dolinski2019,yhkim2020arxiv}.

In thermal calorimetric detection, as discussed in Section~\ref{sec:sensor}, the detector setup mostly consists of   a temperature sensor and an absorber. For  \zerodbd{} experiments based on thermal calorimetric detection, the temperature sensor can be chosen from the high-resolution technologies introduced in Section~\ref{sec:sensor}, while the absorber should be a dielectric crystal containing double beta decaying isotopes.

The sensors should have a high resolution not only for the energy region of \zerodbd{} signals of a few MeV but also for a much wider energy range to investigate background signals.
For this reason, MMCs and NTD Ge thermistors are suitable for the sensor of a \zerodbd{} search experiment because these sensors have a very wide practical dynamic range as well as a well-characterized detector non-linearity, compared to other superconducting sensors such as TESs and KIDs.

For absorber selections, scintillating crystals are preferred because unwanted background signals can be discriminated through the simultaneous detection of light (scintillation) and heat (phonon) signals, as shown in Figure~\ref{fig:particle_id}. Furthermore, these materials have better thermal properties than noncrystalline materials. 

To measure scintillation light for PID, a light detector  composed of a thin Si or Ge wafer can be employed, similarly to the CRESST setup for DM detection. 
Note that the total energy of the scintillation photons absorbed in a light detector is typically on the order of keV in \zerodbd{} search experiments  that is orders of magnitude greater than in direct DM detection experiments mainly because of the difference in the amount of energy absorbed in the scintillating crystal.
For such an energy scale, the light detector of \zerodbd{} search experiments has a large selection of sensor technologies, including those with a relatively small dynamic range such as TESs and KIDs.

Below, we discuss ongoing \zerodbd{} projects in detail that have adopted MMC and NTD Ge as both the heat and light sensors. We also discuss future \zerodbd{} projects that have chosen TESs and KIDs as the light sensors as well as other related techniques used in light detectors.

\color{black}

	\subsubsection{AMoRE} 

~The Advanced Mo-based Rare-event search Experiment (AMoRE) is a large-scale international project to search for \zerodbd{} of \Mo{100} in a deep underground laboratory. 
Similar to the previously discussed DM detection experiments, AMoRE employs superconducting detectors to probe extremely rare \zerodbd{} signals. The AMoRE detectors consist of heat and light detectors to  measure the heat and light signals, respectively, induced by energy input in scintillating crystals, a detection scheme capable of PID analogous to that of CRESST. However, unlike DM detectors, AMoRE detectors are optimized to measure electron-induced events on the MeV energy scale.

AMoRE takes advantage of the high Q-value of \twodbd{} of \Mo{100} (3034\,keV). Because this Q-value is greater than the energy of most environmental $\gamma$-rays, especially the $^{208}$Tl line of 2615\,keV, the background rate in the region of interest (ROI) is significantly reduced. Nevertheless, another source of background due to $\alpha$ decays inside or near the absorber crystal may overwhelm the possible \zerodbd{} signal unless it is properly eliminated. This is why the AMoRE detectors require PID capability for background rejection in addition to a high energy resolution. 

AMoRE has involved tremendous effort in the development of heat and light detectors as well as their target absorber crystals. The heat and light detectors of AMoRE both employ high-resolution MMC sensors. For the absorber, among several candidates, \CMO{} and Li$_2$MoO$_4$ have been chosen based on their high light yield and reasonable cost for mass production. In the first cryogenic test of a small \CMO{} with MMC, reasonable energy resolution in a wide energy range was achieved, demonstrating the applicability of the \CMO{}--MMC combination detector~\cite{sjlee2011}. Soon, a detector with much larger crystals of \CMO{} and Li$_2$MoO$_4$ ($>200$\,g) equipped with a separate light detector was successfully built, which became the basis of the AMoRE detector module~\cite{gbkim2015, gbkim2017app}.

\begin{figure}[t]
    \begin{center}
    \includegraphics[width=8cm, keepaspectratio]{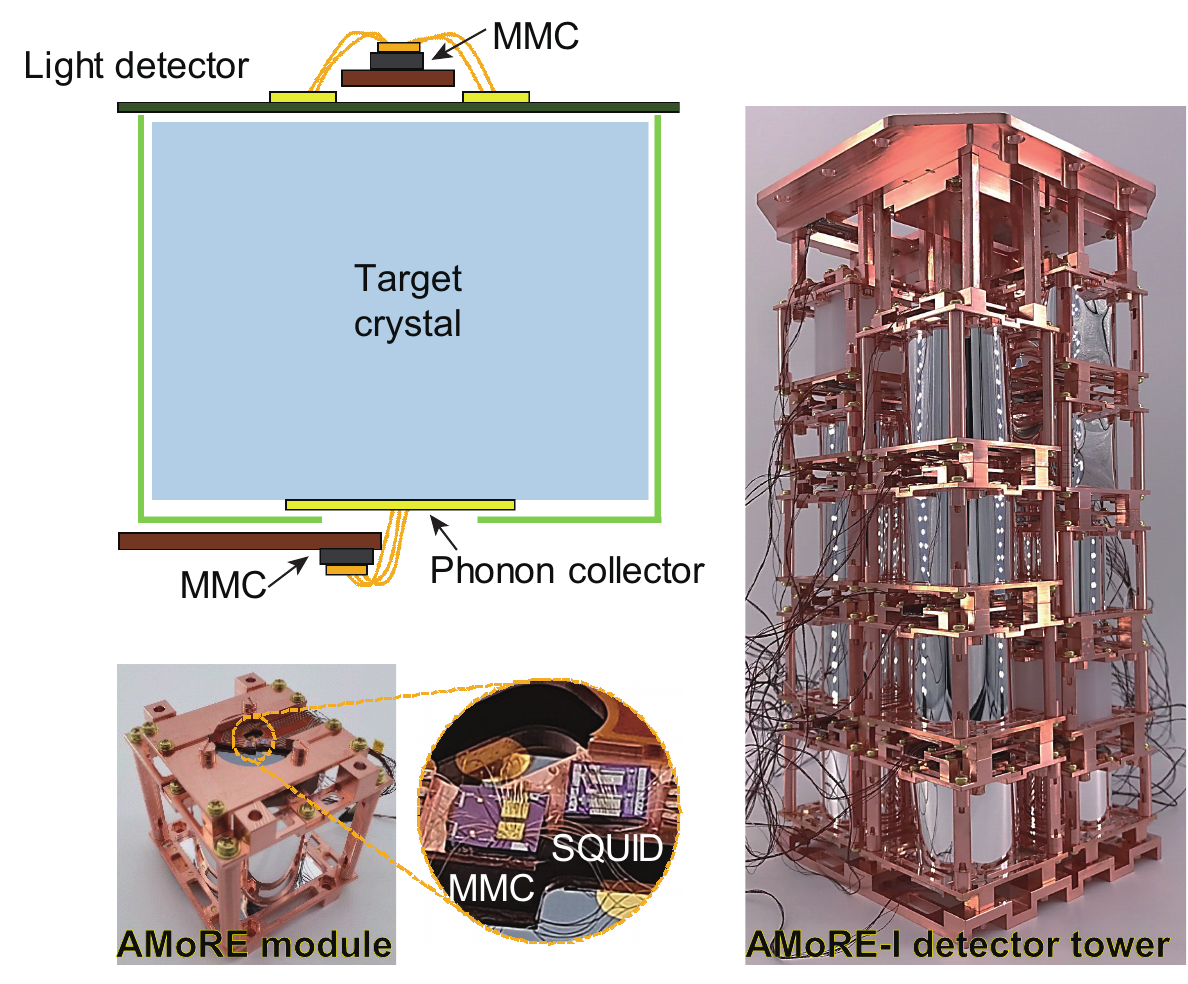}
    \end{center}
    \caption{A simplified schematic ({\it top left}) and a photograph ({\it bottom left}) of an AMoRE module as well as a photograph of the assembled AMoRE-I detector tower with 18 modules ({\it right}).  Each detector module includes one scintillating crystal with two MMC channels for phonon  and scintillation signals. The figure was taken from~\cite{sgkim2021ieeetas}}
    \label{fig:AMoRE_tower}
\end{figure}

The AMoRE detector consists of multiple modules, each consisting of a  target crystal of \CMO{} or \LMO{} and a light detector enclosed by highly light-reflecting material, as shown in Figure~\ref{fig:AMoRE_tower}.
In each module, two MMC sensors are used, one for the heat (phonon) channel and the other for the light channel. To increase the sensitivity of the heat channel, which is AMoRE's main signal, a gold film that serves as a phonon collector is evaporated on one side of an absorber crystal. Then, an MMC sensor is thermally connected to the phonon collector film via gold bonding wires. As discussed in the Section~\ref{sec:sensor}, the input energy in the crystal is transferred to the MMC sensor via athermal and thermal processes of heat flow. The athermal phonons generated in the crystal in the very early stage of energy absorption travel around the crystal and have a significant probability of directly depositing their energy onto the phonon collector before downconversion to thermal phonons. The heat transferred to the phonon collector flows into the MMC sensor and eventually to the heat bath. The light detector is built on the basis of an almost identical heat flow mechanism, with two main differences: a Si or Ge wafer is used as the absorber for lights from the main scintillating crystal, and input energy, deposited by thousands of photons, is absorbed in a broad area than being localized, as is the case for the heat detector.

The AMoRE detectors were carefully designed based on complete thermal modeling that involves Kapitza conductance,  electron-phonon interaction, and electronic heat flow following the Wiedenmann-Franz law~\cite{gbkim2014}.
Thus, optimized AMoRE detectors have shown an energy resolution of $\sim15$\,keV FWHM at 3\,MeV.
Although 15\,keV FWHM is considered a high resolution for a large detector, it can be further improved by reducing the vibration of the cryostat~\cite{lee2018vibration} and by reducing the position dependency~\cite{amore2019epjc}. In the current design, due to the finite thermalization time of the large crystal, the signal size for a given energy input varies depending on the position of the input, thus decreasing the energy resolution. Employing an additional MMC sensor is expected to not only solve the position dependency problem but also provide extra information on the position of the event through the signal size difference between the two sensors. This extra information can be used to distinguish localized \zerodbd{} events from delocalized high-energy $\gamma$-ray events subject to multiple scattering inside the crystal. AMoRE detectors have also demonstrated a fast rise-time of a few ms.  
This fast rise-time is realized by the use of MMCs and gold phonon collectors and is much faster than that of a crystal detector of similar dimensions equipped with an NTD Ge thermistor, a popular choice in several \zerodbd{} experiments. 
A fast detector is very advantageous in cryogenic \zerodbd{} experiments because it reduces unresolved random pileups, which can be one of the most significant background signals in searches for \zerodbd{} signals of \Mo{100} in particular~\cite{gbkim2017app,chernyak2012random}.

Recently, AMoRE finished its pilot stage run, which used 1.9\,kg of $^{\rm 48dep}$Ca$^{100}$MoO$_4$ crystals, depleted in $^{48}$Ca and enriched in \Mo{100}, at Yangyang underground laboratory (Y2L).
\Ca{48} was depleted because \Ca{48} undergoes double beta decay with a Q-value of 4268\,keV, higher than that of \Mo{100}, and it can contribute to the background in the ROI that is indistinguishable from the \Mo{100} signal. The pilot run resulted in a \Tzerov{} sensitivity of 9.5$\times$10$^{22}$\,years~\cite{amore2019epjc}. 
Presently, an upgrade version of AMoRE-I equipped with 6 kg of $^{\rm 48dep}$Ca$^{100}$MoO$_4$ and \enLMO{} just started data collection. Moreover, AMoRE-II has been fully funded, with a plan to build a 200 kg detector of \enLMO{} with a lower cosmic background to achieve a limit sensitivity of \Tzerov{} $>$  5$\times$10$^{26}$ years or, equivalently, \mbb{} $<$ (0.017--0.029)\,eV \cite{amore2019taup}. The experimental site and facility are being prepared for the large-scale superconducting detector of AMoRE-II in Yemilab, a new underground lab in Korea~\cite{park2021taup}.


AMoRE's MMC-based detection scheme has been adopted by another \zerodbd{} experiment: CAlcium fluoride for the study of Neutrinos and Dark matters by Low Energy Spectrometer (CANDLES)~\cite{tetsuno2020jop,li2020jop}. The CANDLES experiment has been searching for \zerodbd{} signals of $^{48}$Ca from CaF$_2$ crystals using liquid scintillators as their main detector.  As a strategic R\&D project, CANDLES recently carried out R\&D experiments in collaboration with AMoRE, successfully demonstrating its PID capability based on simultaneous measurement of heat and light signals from a CaF$_2$ crystal~\cite{tetsuno2020jop,li2020jop}.

	\subsubsection{Other low-temperature thermal calorimeters for \zerodbd{} searches}  

%

Low-temperature thermal calorimetric detectors have long been studied for use in \zerodbd{}   search experiments~\cite{fiorini1984low}. In particular, a series of projects based in Italy have investigated the development of large-mass low-temperature detectors for more than 30 years~\cite{arnaboldi2003app}. Preceded by Milano Double Beta Decay (MiDBD) and CUORICINO, the Cryogenic Underground Observatory for Rare Events (CUORE) project has utilized 800 kg TeO$_2$ crystals as the source/detector of \Te{130} \zerodbd{} events and began collecting data in 2017. All three experiments were conducted in the Laboratori Nazionali del Gran Sasso (LNGS) underground laboratory in Italy~\cite{arnaboldi2003app,gorla2006cuoricino,alduino2018first}. 
A new project called the CUORE Upgrade with Particle Identification (CUPID) is under development as a successor of the CUORE project but with an improved background rejection capability ~\cite{beretta2021appliedsci}. 

The CUORE detectors use NTD Ge thermistors as temperature sensors of the TeO$_2$ crystals.
For each crystal, an NTD Ge thermistor is glued onto the crystal surface. In this kind of setup, the heat flow between the crystal and the sensor at low temperatures is governed by phonon transfer between interfaces of difference media and interactions between phonons and electrons in the sensor, as discussed in Sec.~\ref{subsec:sensor-amm} and \ref{subsec:sensor-ep}. 
Moreover, heat transfer by athermal processes is greatly suppressed.
As a result, the typical rise and decay times of heat signals in the CUORE detectors are relatively slow, on the order of a few hundred milliseconds and several seconds, respectively. 
This inefficient heat flow was intended to function as an efficient low-pass filter of the detector signals because it can help increase the energy resolution of such a massive detector.
Thermistors are a suitable choice for CUORE detectors also because they have a large dynamic range of several MeV.
On the other hand, the slow response of the thermistor makes the detector particularly vulnerable to microphonic noise, such as that caused by mechanical vibration, typically with $1/f$ characteristics. Thus, it is critical to minimize the vibration of the cryostat for CUORE. After all these considerations, the CUORE detectors have demonstrated high energy resolution in their ROI (7.7\,keV FWHM near 2.5\,MeV)~\cite{alduino2018first} .

The CUORICINO and CUORE experiments resulted in the best detection limit for \zerodbd{} decay of \Te{130} to date (a half-life of 3.2$\times$10$^{25}$\,years, which corresponds to the Majorana neutrino mass of 75 $<$ \mbb{} $ <$ 350\,meV~\cite{adams2020prl}). The next-generation experiment CUPID aims not only to employ scintillating crystals to improve the background through PID but also to investigate other isotopes, such as \Se{82} and \Mo{100}. In particular, the CUPID-Mo experiment based on \LMO{} crystals with NTD Ge thermistors has been carried out to demonstrate the applicability of CUPID~\cite{armengaud2021new}.
Moreover, in CUPID-0, the first pilot experiment of CUPID, the PID technique based on heat-light detection resulted in suppression of background signals induced by alpha particles in Zn$^{82}$Se crystals~\cite{azzolini2019prl}. 
Together with AMoRE, CUPID is expected to be a key player in the \zerodbd{} searches for the next decade.

	\subsubsection{Superconducting light detectors for \zerodbd{} experiments} 
%

The PID capability is critical in \zerodbd{} searches to efficiently reject background signals. Although PID using only the heat channel has been demonstrated in a few scintillating crystals~\cite{gbkim2017app,armengaud2020epjc}, in general, simultaneous measurement of heat and light signals using a separate light detector is superior to the single-channel-based PID. A separate light detector may also enable dual-channel-based PID even with a nonscintillating crystal such as TeO$_2$~\cite{cuore_luke} because energetic electrons and ions in a dielectric crystal generate a different amount of Cherenkov light. This capability will broaden the selection of crystals and facilitate searching for \zerodbd{} in multiple isotopes.  Moreover, having a light detector is beneficial not only during \zerodbd{} data acquisition but also during detector development and characterization in the R\&D stages. 
This is why many of the current \zerodbd{} search experiments (e.g., AMoRE, CUPID, and CUPID-Mo) have adopted scintillating crystals and light detectors.
However, the total energy released in the light channel is often very weak. 
For example, only a few percent of the total energy is carried away by scintillation light from scintillating crystals, or in the case of a nonscintillating crystal, Cherenkov photons with a total energy of 100--200\,eV are created by a few MeV electrons absorbed in the nonscintillating crystal~\cite{cuore_luke}. 
Therefore, the light detector should be carefully designed to maximize the size of the light signals. 

The light detector should be implemented with a sensor with high energy resolution and high sensitivity. Superconducting sensors such as TES, MMC, and KID are a good choice~\cite{rothe2018jltp,hjlee2015,casali2019epjc}. These sensors are also advantageous because their intrinsic response time is very fast, and the overall response time can be accelerated by making direct contact between the sensor and the light absorber or by placing a phonon collector film between them, which makes these detectors sensitive to both athermal and thermal phonons. The athermal phonon signal contributes a large portion of the total heat signal in the first few milliseconds and can be utilized to extract extra information that the thermal phonon signal alone cannot reveal.

In addition,
if the light signal is still too weak even with a high-sensitivity sensor, NTL phonon amplification can be applied to the light detectors, especially when the absorber is a Si or Ge wafer, as discussed in Sec.~\ref{subsec:sensor-NTL}. Large-gain signal amplification has been demonstrated with TES~\cite{STARK2005738} and MMC~\cite{jajeon2020jltp} as well as NTD Ge~\cite{cuore_luke} devices.

\subsection{Noncalorimetric approaches for wave-like DM} 

As axions and ALPs emerge as strong  candidates for wave-like DM of sub-eV mass, many detection technologies have been developed to search for these hypothetical particles. 
One powerful approach is to detect excess electromagnetic (photon) signals of a certain frequency converted from axions or ALPs in a strong magnetic field, the so-called inverse Primakoff effect, as illustrated in Figure~\ref{fig:detection_methods}.
Since the possible mass range of wave-like DM spans orders of magnitude, different detection schemes are required depending on the target mass (frequency) of the DM. For the detection of high-frequency DM, it is desirable to use a microwave cavity and quantum-limited amplifiers, which are often composed of superconductors. However, for ultralow-mass (low frequency) DM, the dimension of a microwave cavity that matches the frequency of the excess photons becomes impractically large. In this case, a superconducting lumped element circuit can be adopted to directly sense the electromagnetic signals converted from ultralow-mass DM. Below, we review both microwave cavity-based and lumped circuit-based detection methods.

	\subsubsection{Axion searches with microwave cavities}

\begin{figure}[t]
\begin{center}
\includegraphics[width=7cm]{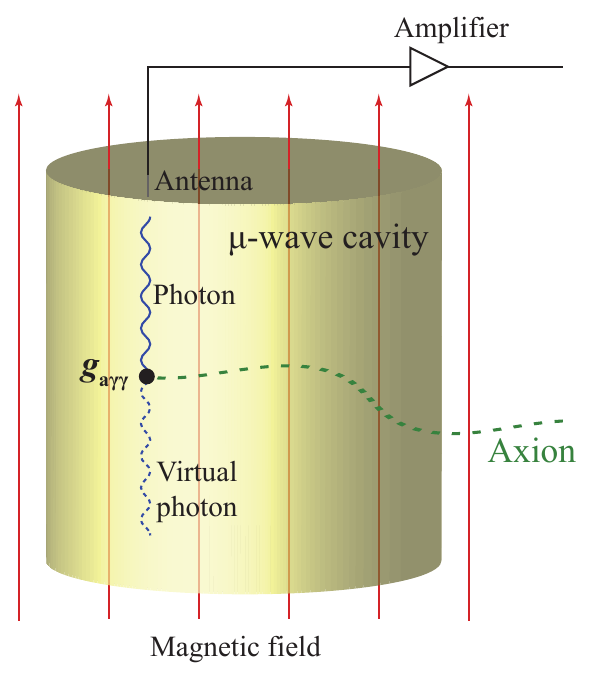}
\end{center}
\caption{Schematic diagram of an axion haloscope based on a microwave cavity. Under a strong magnetic field, an axion interacts with a virtual photon (a quantum fluctuation of the magnetic field) and is converted to a real photon due to the inverse Primakoff effect.
The conversion rate depends on the axion-photon coupling constant $g_{a\gamma\gamma}$. When the frequency of the photon created inside the microwave cavity matches the resonance frequency of the cavity, the detection sensitivity is maximized. Quantum-measurement-based superconducting devices such as SQUIDs and JPAs can be adopted for the frond-end amplifier.}
\label{fig:cavity}
\end{figure}

The Primakoff effect describes the resonant production of neutral pseudoscalar particles by photons interacting with an
atomic nucleus~\cite{Primakoff:1951}.
Its reverse process, the inverse Primakoff effect/scattering, provides an excellent way to detect axions; an axion can decay into two real photons or be converted to a photon through interaction with a virtual photon in an electromagnetic field. The decay rate can be enhanced by a strong magnetic field.

The current most sensitive axion detectors were developed based on the axion haloscope proposed by
Sikivie~\cite{1984:Sikivie}. As shown in Figure~\ref{fig:cavity}, in an axion haloscope,  a microwave cavity is adopted to detect photons produced via the inverse Primakoff effect.
The detection sensitivity of the axion-photon conversion is enhanced when the resonant frequency of a microwave cavity matches the axion mass. The resonant frequency of the cavity is adjustable with a tuning rod.
An antenna placed in the cavity detects the induced microwave power in the cavity and transfers the signal into the cryogenic receiver chain.

Because an extremely low conversion rate is expected, several experimental strategies should be used to increase the signal-to-noise ratio (SNR) of photon detection in an axion haloscope. 
For instance, a microwave cavity can be implemented with a large superconducting magnet to increase the axion-photon conversion rate. 
Moreover, to minimize the overall noise, the temperature of the microwave cavity should be lowered to less than 100\,mK to reduce the thermal noise of the measurement system, and an amplifier with a highest possible gain should be used.
A longer measurement time would also increase the SNR and increase the chance of detecting the rare signal.

Because the major noise in an axion haloscope is thermal noise, the SNR can be described by the Dicke radiometer equation~\cite{1946:Dicke,Stern:2015kzo,Sikivie:2020zpn}:
\begin{equation}
\mathrm{SNR} = \frac{P}{k_\mathrm{B} T_\mathrm{sys}} \sqrt{\frac{t}{b_a}},
\end{equation}
where $P$ is the expected signal power, $k_\mathrm{B}$ is the Boltzmann constant, $T_\mathrm{sys}$ is the system noise temperature,
$t$ is the integration time,
and $b_a$ is the axion signal bandwidth.
For given experimental parameters, the signal power $P$ is expressed as
\begin{equation}
P \propto g_{a\gamma\gamma}^2 B^2 V Q_\mathrm{L},
\end{equation}
where $g_{a\gamma\gamma}$, $B$, $V$,  and $Q_\mathrm{L}$ are the model-dependent axion-photon coupling constant, magnetic field, cavity volume, and loaded quality factor of the cavity, respectively~\cite{Stern:2015kzo}.
The axion signal bandwidth is determined by the energy dispersion of the axion and can be expressed as $b_a=m_{a}v\delta v /c^2 = m_{a}/Q_{a}$, where $m_a$ is the axion mass, $v$ is the axion velocity, $\delta v$ is the velocity dispersion, and $Q_{a}$ is the axion  ``quality factor'' on the order of $10^6$~\cite{Sikivie:2020zpn}.

The number of signal bandwidths that can be simultaneously scanned is $N = Q_{a}/Q_\mathrm{L}$, where $Q_\mathrm{L}$ is assumed to be smaller than $Q_{a}$, and the cavity bandwidth $b_c$ is $Nb_a$.
Then, the time $\Delta t$ required to scan a frequency range $\Delta f$ to achieve a given SNR is determined by
\begin{equation}
 \Delta t = \frac {\Delta f} {b_c} t  \propto  \Delta f  \frac{ \mathrm{SNR}^2 T_\mathrm{sys}^{2}}{{g_{a\gamma\gamma}}^4 B^4 V^2 Q_\mathrm{L} Q_{a} }.
\label{eq:integration_time}
\end{equation}
This equation shows that the scan rate $r = \Delta f / \Delta t$ for a target SNR is proportional to ${g_{a\gamma\gamma}}^{4}$,
or equivalently, the sensitivity of the experiment (the minimum $g_{a\gamma\gamma}$ that can be detected) improves only with $r^{1/4}$,
showing the difficulty of the experiment.
Thus, it is highly desired also to improve other experimental parameters in Eq.~(\ref{eq:integration_time}),
namely, the system noise temperature, the magnetic field strength, the cavity volume, and the quality factor of the cavity.

The system noise temperature $T_\mathrm{sys}$ is  approximately given by the following:
\begin{equation}
T_\mathrm{sys} = T_\mathrm{phy} + T_{A1} + \frac{T_{A2}}{G_1} + \frac{T_{A3}}{G_1 G_2} + ... ,
\label{eq:noise_temp}
\end{equation}
where $T_\mathrm{phy}$ is the physical temperature of the cavity and $T_{Ai}$ and $G_i$ are the noise temperature and the gain of the $i$-th stage amplifier in the receiver chain, respectively~\cite{Boutan:2018uoc}.
The noise from the amplifiers, particularly that from the front-end amplifier, is a major contributor to $T_\mathrm{sys}$.
Thus, for efficient reduction in $T_\mathrm{sys}$, 
a voltage-tunable microstrip SQUID amplifier (MSA) is often employed as a front-end amplifier, whose resonance frequency can be adjusted between 0.1 and 0.8\,GHz with the length of the microstrip and the tuning voltage~\cite{Clarke:2001ftz}.
For higher-frequency measurements, Josephson parametric amplifiers (JPAs) can be used. With a modulation frequency of twice the measuring resonance frequency, phase-preserving amplification can be realized in a wide frequency range up to 10\,GHz~\cite{Alesini:2020vny}.

Currently, most axion haloscope experiments are targeting SNR $\sim$5.
With this SNR, in the case of the Axion Dark Matter eXperiment (ADMX) to be introduced later,
 if $Q_\mathrm{L}$ and $T_\mathrm{sys}$ are 30000 and 0.2\,K, respectively, it would take about 6 (300) days to scan $\Delta f$ of 100\,MHz around 0.74\,GHz and achieve 1 KSVZ (DFSZ) level of sensitivity~\cite{ADMX:2020hay}.
 Thus, in order to reach the DFSZ level sensitivity in a wide frequency range within a reasonable experimental period, the above parameters must be further improved.

The current axion-photon coupling limits of several axion haloscopes are shown in Figure~\ref{fig:ax_limit}.
Note that most of the limits set by haloscopes are in the 0.1--10 GHz range. This is because the sensitive frequency region of a haloscope is primarily determined by the dimension of the cavity, which cannot be made arbitrarily large or small for practical reasons. For example, a larger cavity is required to lower the probing frequency, but the size of the cavity is limited by the magnet. On the other hand, a small cavity is required to probe the high-frequency area, but this could result in an unmanageably long scan time as expected by Eq.~\ref{eq:integration_time}.
Below, we introduce some of the major axion haloscope projects and the challenges for the next step.
\begin{itemize}
\item RBF and UF: \\
In the early period of experimental axion searches,
Rochester-Brookhaven-Fermilab (RBF)~\cite{DePanfilis:1987, wuensch1989results}  and the University of Florida (UF)~\cite{Hagmann:1990} conducted axion searches using axion haloscopes based on microwave cavities.
The RBF experiment probed the frequency range of 1.09--3.93\,GHz with a 10\,L copper cavity, a 6\,T magnet, and a cryogenic amplifier based on a GaAs heterojunction field effect transistor (HFET), also known as a high-electron-mobility transistor (HEMT) with the noise temperature of $\sim$12\,K and the physical temperature of cavity was $\sim$4.4\,K.
The UF experiment probed a similar frequency range with a 7\,L copper cavity and a 7.5\,T magnet.
Using cryogenic HFET amplifiers with the noise temperatures as low as 3\,K and the operating temperature of $\sim$2\,K, the detection sensitivity was improved by more than 10 times compared to that of the RBF experiment. 
Although the $g_{a\gamma\gamma}$ limits obtained by these experiments were far above the values predicted by the KSVZ or DFSZ models, they established the foundation of axion search experiments using axion haloscopes.

\item ADMX: \\
Based on the successful demonstration by the two axion haloscope experiments of RBF and UF, the ADMX collaboration built an axion haloscope with sub-GHz target frequencies using a large 200\,L cavity and a 7.6\,T magnet at Lawrence Livermore National Laboratory (LLNL) in 1995, and  this device operated until 2010~\cite{ADMX:1998pbl,Asztalos:2010ADMX}.
Initially, HFETs with a noise temperature slightly above 2\,K were used as the first-stage amplifier, but later, MSAs, whose noise temperatures were well below 100\,mK, were adopted.
From this first-phase experiment, KSVZ axion-photon couplings between 461 and  860\,MHz (1.9--3.53 $\mu$eV) were excluded at a 90\% confidence level. \\
In 2010, ADMX moved to the Center for Experimental Physics and Astrophysics (CENPA) at the University of Washington and achieved several major detector upgrades~\cite{Du:2018ADMX,Braine:ADMX2020,ADMX:2020hay}.
One of the major upgrades was a new cooling system. In the previous phase,
the ADMX microwave cavity was cooled to approximately 2\,K by a pumped liquid helium cryostat.
The new setup used a dilution refrigerator (DR), which cooled the cavity and the first-stage amplifier to the 100 mK range.
Another major upgrade was the replacement of the first amplifier MSAs with JPAs with a tunable resonance,
which enabled searches for axions in the 675--800\,MHz (2.79--3.31\,$\mu$eV)  range,
extended from the $\sim650$\,MHz ($\sim2.69$\,$\mu$eV) region allowed when using the MSAs.
The sensitivity of the ADMX with a 140\,L cavity in these frequency regions reached the DFSZ level.\\
Recently, the ADMX collaboration has developed a new ``sidecar" cavity to search for axions at higher frequencies (4--7\,GHz) that are inaccessible with the main cavity~\cite{Boutan:2018uoc}. To increase the resonant frequency, the sidecar cavity has a tiny volume of 0.38 L. The sidecar cavity experiment published new limits in this high-frequency region~\cite{Boutan:2018uoc} and is planning to broaden their axion search range with several upgrades in their setup, including the adoption of quantum-limited amplifiers, to be operated in tandem with the main ADMX experiment.

\item HAYSTAC:\\
The Haloscope at Yale Sensitive to Axion CDM (HAYSTAC) branched off of ADMX to probe high-frequency axions; HAYSTAC was initially called ADMX-HF.
They utilize a 2\,L cavity, which is much smaller than ADMX's 140\,L cavity.
Recently, they reported a near-quantum-limited sensitivity of approximately 5.75\,GHz ($\sim23.8$\,$\mu$eV) using a JPA~\cite{zhong2018results} and a DR.
Their first result was near  that of the KSVZ prediction except for ADMX \cite{Brubaker:2017HAYSTAC}.
Recently, the detection sensitivity was further improved by using a quantum squeezed-state receiver (SSR), a new technique that circumvents the quantum limit~\cite{backes2021quantum}. Using the SSR, the scan rate was doubled, and a detection sensitivity almost reaching the KSVZ prediction was achieved at around 4.14 GHz ($\sim17.1$\,$\mu$eV).

\item CAPP:\\
The Center for Axion and Precision Physics (CAPP) at the Institute for Basic Science (IBS) was established in 2013 to search for axions in a wide frequency range of 1--10\,GHz.
Recently, CAPP published results from a series of pilot experiments using three different microwave cavities and HEMTs. Lee et al.\ reported the highest sensitivity limit at around 1.62\,GHz ($\sim6.7$\,$\mu$eV) using a 3.5\,L cavity, a superconducting 8\,T magnet, and a DR~\cite{Lee:2020CAPP8TB}. Jeong et al.\ adopted an interesting cavity design called a multiple-cell cavity, a highly efficient way to probe high-frequency regions with a large volume and a relatively simple setup~\cite{Jeong:2020cwz}. Using a double-cell cavity, a 9\,T magnet and He-3 cryogenic system, they obtained the highest sensitivity limit at around 3.25\,GHz ($\sim13.5$\,$\mu$eV). More recently, Kwon et al. reported promising results in the 2.46--2.75\,GHz range using a setup composed of two half-cavities, one with a volume of 0.59\,L and the other with a volume of 1.12\,L, in the same cryostat with an 8\,T magnet and a DR~\cite{Kwon:2020sav}. Especially at around 2.59\,GHz ($\sim10.7$\,$\mu$eV), the limit of this setup was just above the KSVZ prediction.

There are several ongoing R\&D projects within CAPP to enhance the detection sensitivity. These efforts include employing a much stronger magnet, further advancing the multiple-cell cavity approach, and developing a superconducting cavity and quantum-limited amplifiers~\cite{semertzidis2019axion}.
For the stronger magnet, they have been testing a new 12\,T low-temperature superconducting (LTS) magnet with a bore size of 32\,cm and an 18\,T high-temperature superconducting (HTS) magnet with a bore size of 7\,cm~\cite{Kim:2020jaw}.
Although their 18\,T magnet is the strongest magnet ever adopted for axion haloscopes, its relatively small bore limits the size of the cavity. They plan to upgrade the magnet to a 25\,T HTS magnet with a 10\,cm bore that is under development~\cite{KimJG:2019abc,Gupta:2019abc}. 
For the superconducting cavity, they demonstrated a cavity quality factor 6 times higher than that of their copper cavity using a polygon-shaped cavity with commercial YBCO tapes covering the entire inner wall~\cite{ahn2020superconducting}.
For the quantum-limited amplifiers, R\&D on JPAs is ongoing~\cite{kutlu2021characterization}. Employing JPAs as first-stage amplifiers instead of HEMTs will decrease the noise temperature to under 1\,K and improve the sensitivity.
With all the planned upgrades, CAPP is expected to reach the DFSZ limit in the 0.7--3\,GHz range using the 12\,T magnet and the KSVZ limit at frequencies up to 10\,GHz using the 25\,T magnet~\cite{semertzidis2019axion}.

\item QUAX:\\
QUaerere AXion, or QUest for AXion, (QUAX) is a collaboration searching for axion or ALPs in two different approaches, one based on the axion-photon coupling (the QUAX-$a\gamma$ experiment) and the other on the axion-electron coupling (the QUAX-$ae$ experiment~\cite{Barbieri:2016vwg}). Here, we focus only on the former. 

In 2019, they reported the first axion-photon coupling limit around 37\,$\mu$eV ($\sim$9\,GHz) using a 36.43\,mL NbTi superconducting cavity with a high $Q_\mathrm{L}$ of 2.0$\times$10$^5$ and a magnetic field of 2\,T~\cite{Alesini:2019ajt}.
It was the first axion limit result from an axion haloscope built on a superconducting cavity.
In 2021, they obtained an improved axion limit close to KSVZ around 43.0\,$\mu$eV (10.4\,GHz) using a 80.56\,mL cavity made of Cu with $Q_\mathrm{L}$ of 36000 and 8.1\,T at an operating temperature of 200\,mK~\cite{Alesini:2020vny}. The improved limit was also attributed to lowering the noise temperature from 15.3\,K to 0.9\,K using a JPA.  

QUAX-$a\gamma$ is currently building two detectors, one in the Laboratori Nazionali di Legnaro (LNL) and the other in the Laboratori Nazionali di Frascati (LNF). 
They are complementary detectors built upon slightly different techniques. The LNL detector will be based on a cavity made of hollow dielectric cylinders covering 10--11\,GHz~\cite{QUAX:2020uxy} while the LNF detector will be based on a multicavity with the cavities tuned to different frequencies, covering 9--10\,GHz~\cite{universe7070236}.

\end{itemize}

	\subsubsection{Direct detection with superconducting lumped element circuits}	


For the search for axions and ALPs in the mass range below 1\,$\mu$eV (or 0.2\,GHz), the use of microwave cavities is not practical and sometimes even impossible because the dimensions of the cavities with such a low resonance frequency will exceed that of any available strong magnets. 
Instead, a lumped superconducting circuit can be employed to directly sense the electromagnetic signal induced by the photons converted from axions via the inverse Primakoff effect.

\begin{figure}[t] 
\begin{center}
\includegraphics[width=8cm]{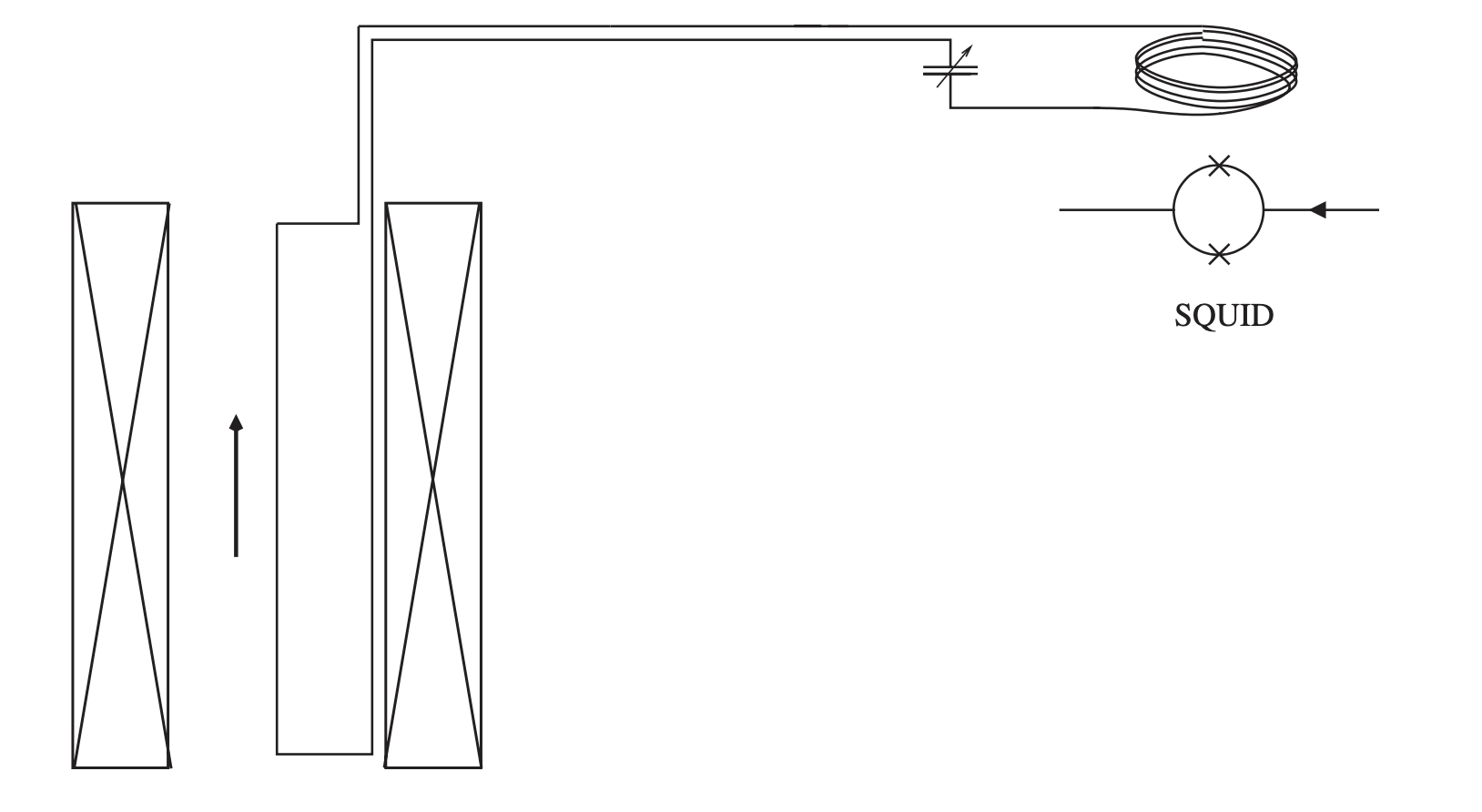}
\end{center}
\caption{A simplified measurement circuit for axion DM using a superconducting LC circuit. The small coil of the circuit is coupled to a dc-SQUID. The figure was taken  from \cite{sikivie2014prl}.
}
\label{fig:lc_circuit}       
\end{figure}

Figure~\ref{fig:lc_circuit} describes an axion detection scheme using a resonator circuit. A superconducting pickup loop is placed in a strong magnetic field to boost axion-photon conversion.
This scheme overcomes the practical limit of placing a large microwave cavity inside the bore of a fixed-size high-field magnet to search for low-mass (frequency) axions. Note that the maximum magnetic field strength in these experiments is well below the critical magnetic field of the superconducting wire typically made of Nb-Ti alloy ($\sim$15\,T).
The other end of the pickup loop is coupled to a magnetometer such as a SQUID.
Since the coupling constant $g_{a\gamma \gamma}$ predicted by the  KSVZ and DFSZ models becomes weaker as the axion mass becomes lighter, the magnetometer should be extremely sensitive for ultralow mass axion searches.
The technological maturity of superconducting sensors and superconducting electronics may provide the required sensitivity over a wide frequency region between DC and a few hundred MHz. 

We introduce three projects searching for low-mass axion DM based on superconducting circuits and superconducting sensor technologies. 

\begin{itemize}
\item ABRACADABRA: \\
A Broadband/Resonant Approach to Cosmic Axion Detection with an Amplifying B-field Ring Apparatus (ABRACADABRA) uses a toroidal magnet that induces a large circular magnetic field and a superconducting pickup loop placed in the center of the toroidal structure. 
The pickup coil is meant to measure electromagnetic waves along the circular axial volume due to axion-photon conversion under the magnetic field. 
A sensitive dc-SQUID is inductively coupled to the pickup loop with a superconducting flux transformer to match the inductance of the SQUID input coil and the pickup loop.  The detector apparatus also includes a current-injection coil along the axial line of the toroidal coil for calibration purposes.   
Recently, ABRACADABRA-10\,cm, a prototype detector, was built with an active volume of 890\,cm$^3$ and a 10\,cm 1\,T toroidal magnet, and its performance was tested. Broad-band detection was carried out  in the $m_a$ range of 0.3--8.1\,neV, resulting in a $g_{a\gamma \gamma}$ sensitivity as low as approximately 10$^{-10}$\,GeV$^{-1}$~\cite{ouellet2019prl}.

\item ADMX SLIC: \\
ADMX Superconducting LC Circuit Investigating Cold Axions (ADMX SLIC) is another project using a superconducting lumped element circuit for axion DM searches. The idea was proposed in Ref.~\cite{sikivie2014prl}, and the first measurement in this project was carried out recently~\cite{crisosto2020prl}. 
A superconducting LC circuit was introduced into an ADMX-type 7\,T magnet, which is a cylindrical magnet with a large bore, as shown in Fig.~\ref{fig:lc_circuit}. 
A rectangular loop antenna was placed inside the magnet to measure the electromagnetic signals induced by the photons converted from axions.
The antenna was made of a copper-matrix-free 0.25\,mm
diameter NbTi wire strung around a polytetrafluoroethylene (PTFE) structure of  7.62\,cm $\times$ 31.25\,cm.
The LC resonance circuit was equipped with a piezoelectric-driven capacitor placed at the low-temperature stage to be able to tune the resonance frequency of the circuit. 
The first measurement resulted in $g_{a \gamma \gamma}$ sensitivities below 10$^{-12}$\,GeV$^{-1}$ in the narrow ranges of  (1.7498--1.7519) $\times$ 10$^{-7}$ eV, (1.7734--1.7738) $\times$ 10$^{-7}$\,eV, and (1.8007--1.8015) $\times$ 10$^{-7}$\,eV~\cite{crisosto2020prl}.  

\item DM Radio: \\
Dark Matter Radio (DM Radio) is another DM detector based on a tunable lumped-element LC resonator~\cite{silva2016IEEEtas}. 
The inductor is a superconducting coil wrapped around a pickup sheath composed of Nb. The capacitor is tunable using movable sapphire dielectrics placed inside a Nb capacitor. The excess power induced by axion-photon conversion above the thermal noise level is sensed by dc-SQUIDs, ac-SQUIDs, or parametric amplifiers depending on the target frequency.
The overall detection scheme is similar to that in other projects based on a superconducting LC circuit, but the resonator of DM Radio is enclosed in a Nb superconducting shield, which keeps the active detector volume free from electromagnetic noise from the surroundings while having no effect on the axions. This setup allows DM Radio to detect axions as well as ultralight hidden photons (or dark photons). A hidden photon is a hypothetical particle that interacts with normal photons via kinetic mixing with an extremely small mixing angle $\epsilon$~\cite{graham2016vector}.  Hidden photons have many phenomenological similarities with axions, but unlike axions, their detection does not require a strong magnetic field.

A proof-of-principle detector of 0.1\,L was designed with an LC circuit with a fixed resonance frequency of 492\,kHz and a quality factor of $\sim40000$. Since this detector did not have a magnet for axion-photon conversion, it was not sensitive to axions but only to hidden photons. A test run with a total integration time of 5.14 hours resulted in an upper limit on the hidden photon mixing angle $\epsilon$ of $\sim$ 10$^{-9}$ at around 2 neV~\cite{phipps2020}. As a next step, a small-scale ($\sim 0.7$\,L) detector called DM Radio-Pathfinder was built with a tunable LC resonator with a quality factor of 148000~\cite{chaudhuri2019dark}. A 1-year scan of the DM Radio-Pathfinder over the 100\,kHz to 10\,MHz range is expected to set an upper limit on $\epsilon$ far below $10^{-10}$~\cite{phipps2020}.

Based on the success of the pilot projects of  ABRACADABRA and DM Radio, the two projects are collaboratively working on a new DM Radio-50L detector. The new detector will be operated under a magnetic field in a dilution refrigerator. After a 1-year scan with a resonator quality factor of $10^6$ and a 0.5\,T magnetic field over a wide range between 40\,peV and 40\,neV (10\,kHz--10\,MHz), the projected sensitivities for axions and hidden photons are $g_{a \gamma \gamma}  \sim 1\times 10^{-15}$\,GeV$^{-1}$ and $\epsilon = 10^{-15}$--$10^{-12}$, respectively~\cite{chaudhuri2019dark}. A future generation detector, DM Radio-M3, with a 1\,m$^3$ volume and a 4\,T magnet is also proposed. The projected axion sensitivity after a 3.5-year scan may reach the KSVZ prediction above 3\,MHz and the DFSZ prediction above 10\,MHz~\cite{chaudhuri2019dark}.  

\end{itemize}

\section{Conclusion}
\label{sec:con}

The two open tasks of the direct detection of DM and the search for double beta decay in neutrinoless mode are among the most important and immediate experiments that deal with fundamental physics questions but have not received plausible answers. 
Superconducting detectors are a type of detector with extremely high sensitivity that have the potential to lead to groundbreaking discoveries related to these tasks.
These novel detectors are based on the measurement of low-temperature quantum phenomena such as superconducting transitions, quasiparticles, or spin dynamics.
These detectors also rely on state-of-the-art superconducting electronics such as SQUIDs and parametric amplifiers to improve their signal-to-noise ratio, which is essential in rare event search experiments.
Many international experiments have been carried out recently and are planned to be built with the superconducting detector technologies.
Many international experiments have been carried out recently and are planned to be conducted with superconducting detector technologies.

Direct DM detection experiments can be divided into two groups depending on the mass of the DM particles, which determines the type of detector technology to be used. For particle-like DM heavier than $\sim$1 eV, superconducting detectors look for a signal induced by the absorption or scattering of DM particles in a target material. 
To date, the main goal of this approach has been increasing the total mass of the target material and the energy resolution of the detector. 
However, the focus is shifting toward the detection of low-mass DM particles by lowering the detection energy threshold. For much lighter (wave-like) DM particles, low-temperature quantum amplifiers are adopted, where the wave-like DM particles are converted to photons or induced in a microwave cavity or superconducting circuit. 

In the search for the \zerodbd{} process, the energy resolution and  timing resolution are important parameters for a large-scale low-temperature experiment with crystal targets. In these experiments, superconducting detectors are often equipped with two different sensors to provide dual-channel detection of phonon and scintillation signals, an essential feature to select the \zerodbd{} signals from unwanted background signals. 

In this review, we discussed the physics of the superconducting detectors used in DM detection and \zerodbd{} search experiments and provided brief introductions to the history, present status, and future plans of several selected experiments.  
This review should be used not only to understand and utilize the technologies specific to the superconducting detectors but also to inspire the development of other  detection methods based on newly developed novel characteristics and phenomena at low temperatures; the existing superconducting detector technologies might be an initiative bridge 
to the desired break-through discoveries. 

We would like to conclude with several examples of other experimental topics in the field of astroparticle physics in which superconducting detectors have played and will continue to play a major role. Several coherent elastic neutrino-nucleus scattering (CE$\nu$NS) experiments have adopted TESs as their main detection technology, leveraging their experience in developing detectors for direct DM search experiments~\cite{billard2017jpg,rothe2020jltp}. In the direct measurement of neutrino mass through electron capture, two active projects, namely, Electron Capture $^{163}$Holmium experiment (ECHo)~\cite{gastaldo2014jltp} and HOLMES~\cite{alpert201epjc}, have adopted MMCs and TESs, respectively. In the search for keV-scale sterile neutrinos as DM candidates, the Beryllium Electron capture in Superconducting Tunnel junctions (BeEST) experiment has adopted superconducting tunnel junction (STJ) detectors~\cite{friedrich2021prl}, another type of superconducting detector with a less superior energy resolution but much faster response than TESs and MMCs in general. Finally, MMCs have also been also proposed as next-generation detectors of axion or ALP search experiments in the keV range~\cite{unger202jinst}.


\ack
Authors are grateful to I. Kim, B.R. Ko, H.S. Lee, S.H. Seo, and S.L. Olsen for valuable discussions. This work was supported by Grant no. IBS-R016-A2.

\section*{References}

\providecommand{\newblock}{}

\end{document}